\documentclass[letterpaper]{article} 
\usepackage[preprint]{aaai2027}  
\usepackage[hyphens]{url}  
\usepackage{graphicx} 
\usepackage{natbib}  
\usepackage{caption} 
\usepackage{algorithm}
\usepackage{algorithmic}
\usepackage{booktabs}      
\usepackage{amssymb}       
\usepackage{amsmath}       
\usepackage{amsmath, amssymb}
\usepackage{multirow}
\usepackage{xcolor}
\usepackage{graphicx}
\usepackage{listings}

\definecolor{starcolor}{HTML}{FFC107}

\usepackage{tikz}

\definecolor{starcolor}{HTML}{FFC107}
\definecolor{starcolor}{HTML}{FFC107}

\newcommand{\goldstar}{%
    \raisebox{0.15ex}{%
        \makebox[1em][c]{%
            \makebox[0pt][c]{\scalebox{1.0}{$\bigstar$}}%
            \makebox[0pt][c]{\scalebox{0.72}{$\textcolor{starcolor}{\bigstar}$}}%
        }%
    }%
}

\usepackage{newfloat}
\usepackage{listings}
\DeclareCaptionStyle{ruled}{labelfont=normalfont,labelsep=colon,strut=off} 
\floatstyle{ruled}
\newfloat{listing}{tb}{lst}{}
\floatname{listing}{Listing}

\usepackage{booktabs}
\title{AudioScape-TTA: A Structured Soundscape Benchmark for Fine-Grained Text-to-Audio Evaluation}
\author{
    Jinting Wang\textsuperscript{\rm 1},
    Yuguang Yang\textsuperscript{\rm 2},
    Shengyu Li\textsuperscript{\rm 1},
    Yan Rong\textsuperscript{\rm 1},\\
    Shan Yang\textsuperscript{\rm 2},
    Xiaoda Yang\textsuperscript{\rm 3},
    Li Liu\textsuperscript{\rm 1}
}
\affiliations{
    \textsuperscript{\rm 1}The Hong Kong University of Science and Technology (Guangzhou)\\
    \textsuperscript{\rm 2}Tencent\\
    \textsuperscript{\rm 3}Zhejiang University
}

\begin{document}

\maketitle

\begin{figure*}[t]
    \centering
    \includegraphics[width=0.99\linewidth]{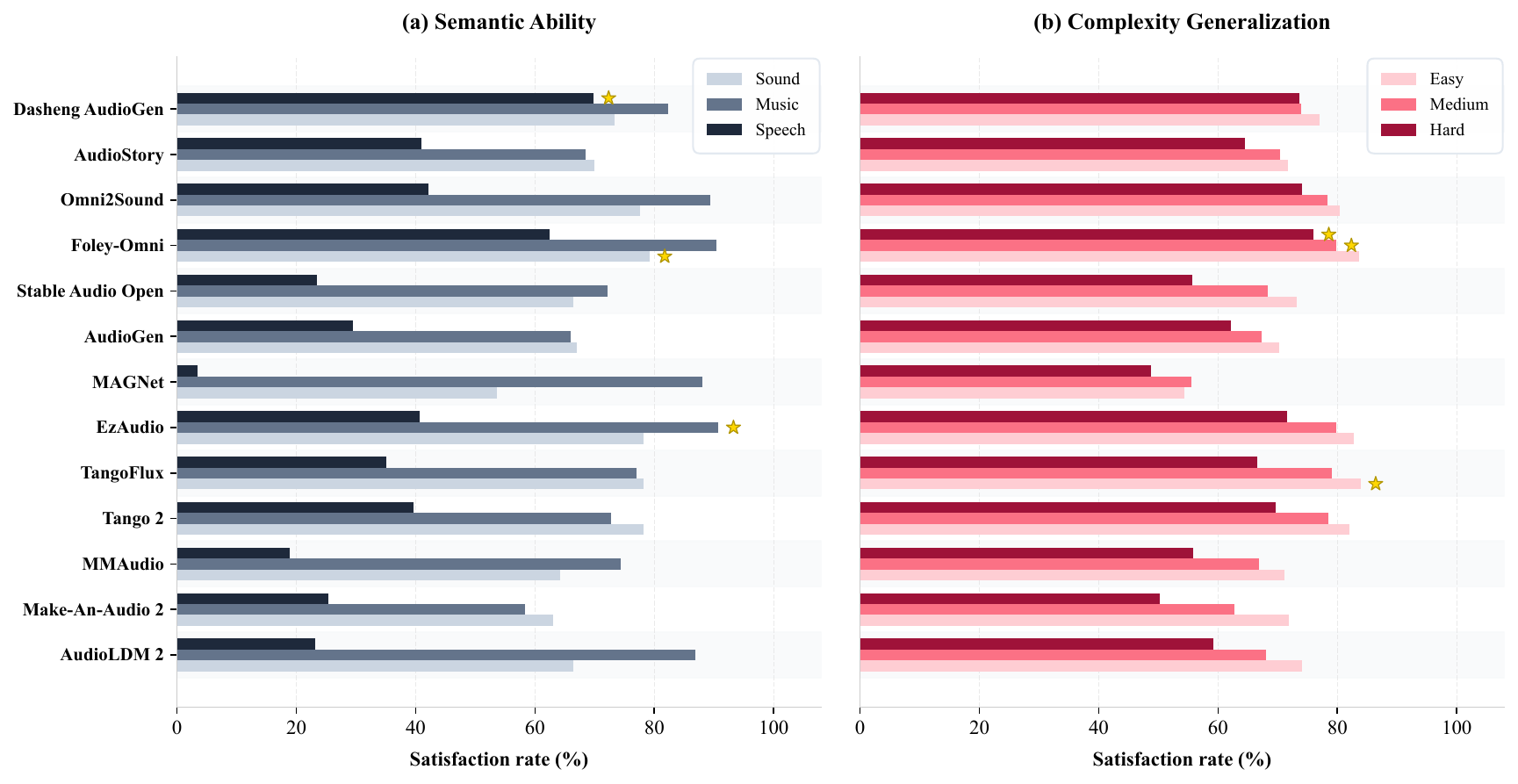}
\caption{Comprehensive evaluation of open-source TTA generation models on the AudioScape-TTA benchmark.
(a) Semantic ability evaluation across three audio modalities, including sound, music, and speech generation.
Speech satisfaction rate includes both speaker-related attributes and content correctness.
(b) Complexity generalization performance under different generation complexity levels, including easy, medium, and hard subsets.
The \goldstar\ symbol denotes the best-performing model for each metric.}
    \label{fig:modality_difficulty}
\end{figure*}

\begin{abstract}
Text-to-audio (TTA) generation has recently achieved remarkable progress in
synthesizing realistic audio from natural language descriptions. However,
determining whether generated audio faithfully satisfies complex textual
instructions remains challenging. Existing benchmarks mainly rely on global
similarity metrics, providing limited insight into fine-grained semantic
failures. To address this limitation, we introduce \textbf{AudioScape-TTA}, a
structured and complexity-aware benchmark for fine-grained TTA evaluation.
AudioScape-TTA represents realistic soundscapes through modality-aware
semantic structures and characterizes generation complexity using event
density and structural complexity. Based on these annotations, we propose a
rubric-based audio-grounded evaluation framework that verifies event
realization, acoustic attributes, and speech content through fine-grained
semantic criteria. The benchmark contains 2,258 audio-text pairs with 25,707
binary QA rubrics, enabling scalable and interpretable analysis of TTA
systems. Experiments on 13 representative open-source TTA models reveal
persistent limitations in fine-grained attribute control, speech-content
preservation, and compositional soundscape generation. Human validation
further demonstrates that our rubric-based evaluation achieves stronger
alignment with human semantic judgments than conventional global similarity
metrics.
\end{abstract}


\section{Introduction}

Recent advances in text-to-audio (TTA) generation have enabled models to synthesize diverse audio components, including sound effects, music, and speech, from textual instructions~\cite{vyas2023audiobox,yang2023uniaudio}. 
With improved generation quality and controllability, TTA systems are increasingly expected to handle complex instructions that describe multiple interacting audio components rather than isolated sound events. 
Therefore, the evaluation challenge is shifting from measuring general audio plausibility toward determining whether models can faithfully realize realistic \emph{structured soundscapes}.

However, existing TTA benchmarks provide limited support for evaluating structured soundscape generation. First, existing benchmarks do not consistently provide realistic and compositional instructions that explicitly represent heterogeneous audio components within a single scene. As summarized in Table~\ref{tab:benchmark_comparison}, prior studies investigate caption-level generation, conditional generation, temporal control, and event-level relations~\cite{kim2019audiocaps,drossos2020clotho,xie2025audiotime,he2025ritta,wang2026tta,wang2025audioatlas}. However, their prompts are largely centered on coarse-grained sound events, offering limited support for jointly specifying sound effects, background music, and speech within a unified soundscape.
In particular, speech is often treated as an audio event rather than a linguistic modality with target utterance constraints, limiting the evaluation of speech-content preservation in complex soundscapes.

Second, existing benchmarks provide limited complexity-aware analysis of structured soundscape generation. Although recent studies consider multi-event prompts and event-level relations, evaluation is typically summarized as aggregate performance. Such evaluation does not characterize sample-level complexity, which may vary with acoustic content, active audio categories, and their clip-level composition. Consequently, existing benchmarks provide limited insight into how semantic instruction following varies as soundscape complexity increases.

Beyond benchmark construction, reliable automatic evaluation remains challenging. Existing metrics primarily rely on global text–audio similarity, such as CLAP~\cite{wu2023large}, or holistic judgments from powerful audio-language models. While these approaches provide useful overall assessments, they typically do not explicitly verify whether individual semantic requirements in structured soundscape instructions are satisfied. Consequently, generated audio may receive favorable global scores while still missing specific components. This motivates a fine-grained evaluation system that provides interpretable diagnoses of semantic failures.

To address these limitations, we introduce \textbf{AudioScape-TTA}, a structured and complexity-aware benchmark for fine-grained TTA evaluation. AudioScape-TTA contains 2,258 instances and 25,707 semantic rubrics. 
Each sample is annotated as a structured soundscape comprising scene context, sound effects, background music, and speech components, together with fine-grained semantic rubrics for semantic verification.
We characterize sample complexity through two complementary dimensions: \emph{event density}, which measures the amount of acoustic content in a clip, and \emph{structural complexity}, which characterizes the diversity and clip-level composition of active audio categories.
 This design enables analysis of model robustness across soundscapes with different levels of acoustic density and compositional complexity.

Building upon this benchmark, we propose an audio-grounded rubrics-based evaluation framework that decomposes each structured soundscape instruction into fine-grained semantic requirements covering event presence, acoustic attributes, and speech content. The framework evaluates these requirements through audio-language understanding and transcription-based verification, providing interpretable diagnostics beyond global text--audio similarity metrics. It further supports fine-grained analysis across audio modalities, soundscape compositions, and complexity levels.

We evaluate 13 open-source TTA models spanning diverse generation paradigms. Results show that global similarity and perceptual-quality metrics provide complementary but incomplete indications of fine-grained semantic following ability. Complexity-aware analysis reveals robustness differences across soundscapes with varying complexity. At the model level, the proposed rubrics-based evaluation achieves stronger correlation with human semantic judgments than CLAP-based similarity, demonstrating the effectiveness of structured semantic assessment.

Our contributions are summarized as follows:
\begin{itemize}
    \item We introduce \textbf{AudioScape-TTA}, a structured and complexity-aware benchmark that represents realistic audio scenes as structured soundscapes and enables fine-grained evaluation across different complexity.

    \item We propose a rubric-based audio-grounded evaluation that decomposes semantic fidelity into event presence, acoustic attributes, and speech content, enabling scalable and interpretable evaluation.

    \item We conduct a comprehensive evaluation of 13 open-source TTA models, revealing key limitations in fine-grained semantic instruction following.
\end{itemize}

\begin{table*}[t]
\centering
\small
\setlength{\tabcolsep}{5pt}
\begin{tabular}{lccccc}
\toprule
\textbf{Benchmark} &
\shortstack{\textbf{Samples}} &
\shortstack{\textbf{Structured}\\\textbf{Schema}} &
\shortstack{\textbf{Speech}\\\textbf{Content}} &
\shortstack{\textbf{Semantic}\\\textbf{Diagnosis}} &
\shortstack{\textbf{Complexity}\\\textbf{Aware}} \\
\midrule

AudioCaps 1.0 \cite{kim2019audiocaps}   
& 975   
& $\times$ 
& $\times$ 
& $\times$
& $\times$ \\

AudioCaps 2.0 \cite{kim2019audiocaps}   
& 1,778 
& $\times$
& $\times$
& $\times$
& $\times$ \\

Clotho  \cite{drossos2020clotho}         
& 1,045 
& $\times$
& $\times$
& $\times$
& $\times$ \\

AudioCondition \cite{guo2024audio}  
& 1,110 
& $\triangle$
& $\times$
& $\triangle$
& $\times$ \\

AudioTime \cite{xie2025audiotime}       
& 500   
& $\triangle$
& $\times$
& $\triangle$
& $\triangle$ \\

RiTTA \cite{he2025ritta}
& 720   
& $\triangle$
& $\times$
& $\triangle$
& $\triangle$ \\

AudioAtlas \cite{wang2025audioatlas}      
& 1,165 
& $\triangle$
& $\times$
& $\triangle$
& $\times$ \\

VinTAGe-Bench \cite{kushwaha2025vintage}   
& 636   
& $\triangle$
& $\times$
& $\triangle$
& $\times$ \\

TTA-Bench \cite{wang2026tta}       
& 2,999 
& $\triangle$
& $\times$
& $\triangle$
& $\triangle$ \\

T2A-EpicBench \cite{wang2025t2a}& 100&$\times$&$\times$&$\triangle$& $\times$ \\

\midrule

\textbf{AudioScape-TTA (Ours)}
& \textbf{2,258}
& \textbf{\checkmark}
& \textbf{\checkmark}
& \textbf{\checkmark}
& \textbf{\checkmark}

\\
\bottomrule
\end{tabular}

\caption{
Comparison of existing TTA benchmarks.
\textit{Structured Schema} indicates whether audio scenes are represented using
predefined components (e.g., sound effects, background music, and speech) rather
than only free-form captions.
\textit{Speech Content} indicates whether synthesized speech is evaluated against
annotated linguistic targets, such as transcripts or specified utterances.
\textit{Semantic Diagnosis} indicates whether benchmarks provide
annotation-grounded criteria for diagnosing specific semantic requirements.
$\checkmark$ denotes full support, while $\triangle$ denotes partial support,
such as event-level, segment-level, or attribute-level evaluation without a unified
semantic rubric.
\textit{Complexity Aware} indicates whether evaluation explicitly characterizes
sample complexity or analyzes performance across different complexity levels.
}
\label{tab:benchmark_comparison}
\end{table*}


\begin{figure*}[t]
    \centering
    \includegraphics[width=0.9\linewidth]{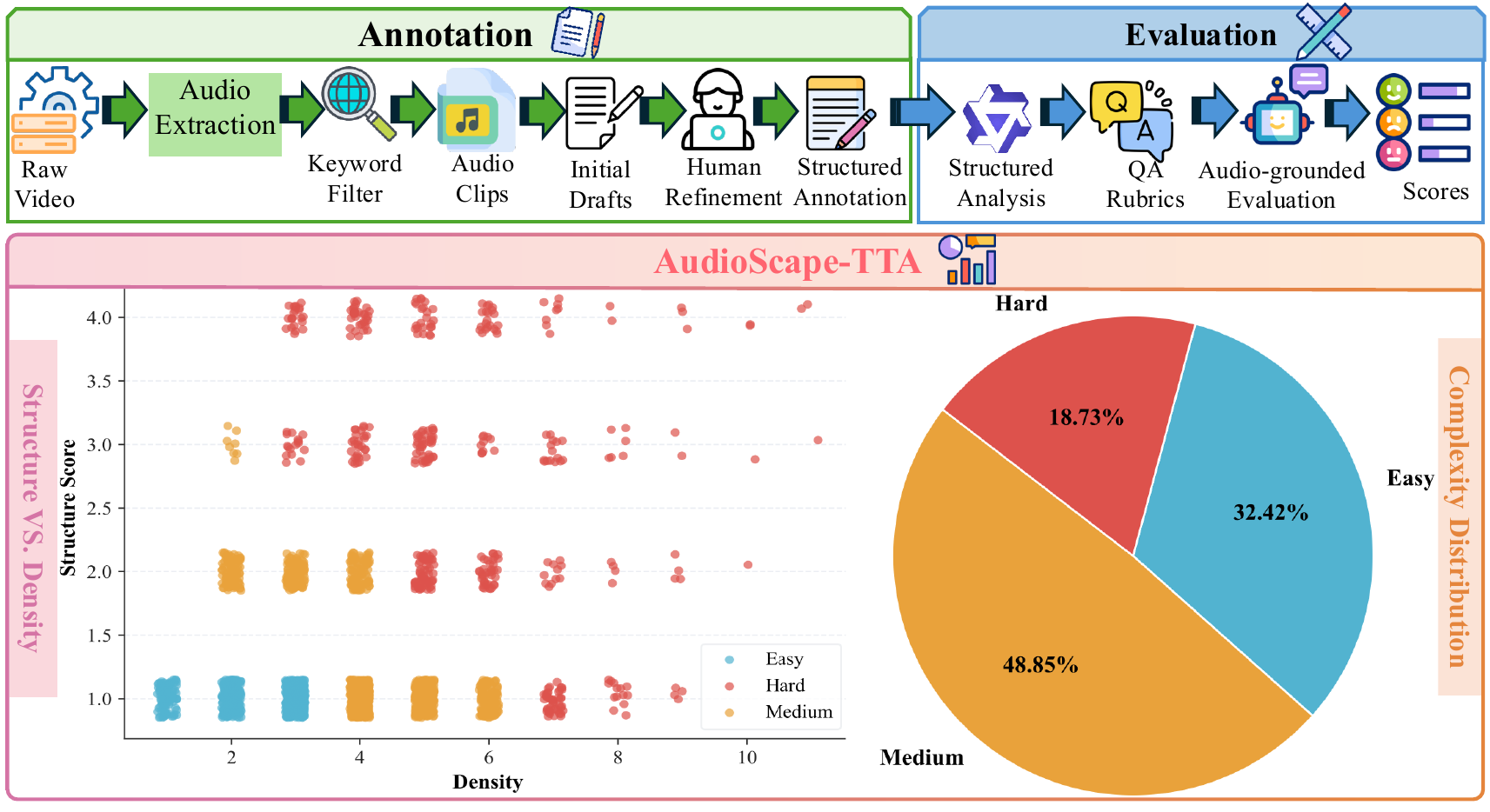}
    \caption{\textbf{Overview of AudioScape-TTA}, a complexity-aware benchmark for text-driven omni-audio generation. The top panel illustrates the benchmark construction and evaluation pipeline. The bottom panel characterizes AudioScape-TTA in a complexity-aware manner by organizing samples within the proposed Density–Structure complexity space.}
\label{fig:overview}
\end{figure*}

\section{Related Work}
\subsection{Text-to-Audio Generation}

Text-to-audio (TTA) generation aims to synthesize realistic audio from
natural-language descriptions. Early approaches, such as AudioGen~\cite{kreuk2022audiogen},
formulate TTA as autoregressive generation over discrete audio tokens,
demonstrating the feasibility of large-scale audio synthesis. Recent progress
has been largely driven by diffusion-based models. AudioLDM~\cite{liu2023audioldm}
introduces latent diffusion for audio generation, while subsequent approaches,
including Make-An-Audio~\cite{huang2023make}, Make-An-Audio~2~\cite{huang2023make2},
Auffusion~\cite{xue2024auffusion}, and EZAudio~\cite{hai2025ezaudio}, further
improve semantic alignment, temporal modeling, and generation efficiency.
More recently, flow-based methods such as TangoFlux~\cite{hung2024tangoflux}
and MeanAudio~\cite{li2026meanaudio} achieve faster generation with competitive
quality. In addition, unified audio generation frameworks, including
Foley-Omni~\cite{tao2026foley}, Omni2Sound~\cite{dai2026omni2sound}, and
AudioStory~\cite{guo2026audiostory}, extend TTA toward multimodal audio synthesis.

Despite recent progress, TTA evaluation remains dominated by perceptual quality
and global text--audio alignment metrics, while fine-grained semantic
requirement satisfaction in structured soundscapes is largely unexplored.

\subsection{Benchmarks and Evaluation of TTA Generation.}

Existing TTA benchmarks evaluate generation capabilities from different
perspectives. Early datasets such as AudioCaps~\cite{kim2019audiocaps} and
Clotho~\cite{drossos2020clotho} provide large-scale audio-caption pairs and
mainly assess global audio--text semantic alignment. However, their free-form
captions do not explicitly represent structured soundscape compositions or
separate heterogeneous audio components.
Recent benchmarks investigate more specific generation capabilities.
AudioCondition~\cite{guo2024audio} explores condition-oriented audio generation,
while AudioTime~\cite{xie2025audiotime} and RiTTA~\cite{he2025ritta} focus on
temporal control and event-level relational reasoning.
TTA-Bench~\cite{wang2026tta} provides a broader evaluation suite covering
diverse prompt categories. AudioAtlas~\cite{wang2025audioatlas} and
VinTAGe-Bench~\cite{kushwaha2025vintage} further extend evaluation toward
realistic movie- and video-oriented audio scenarios.
Recent work has also investigated automatic evaluation of TTA systems through
learned evaluation metrics, preference-based benchmarks, and rubric-based
evaluators, such as AQAScore \cite{kuan2026aqascore}, T2A-EpicBench \cite{wang2025t2a}, and
AnyAudio-Judge \cite{li2026anyaudio}. These approaches primarily focus on improving evaluation
models or measuring overall instruction-following performance, rather than
establishing a standardized benchmark with structured semantic annotations.
In contrast, AudioScape-TTA focuses on benchmark construction. It introduces a
structured soundscape, derives standardized audio-grounded semantic rubrics from hierarchical
annotations, and provides a unified, complexity-aware evaluation protocol for
the reproducible assessment of compositional text-to-audio generation.

\section{AudioScape-TTA}

 \subsection{Overview}

AudioScape-TTA is a structured benchmark and evaluation system designed to
assess whether generated audio faithfully satisfies complex textual
instructions. Instead of relying only on global text--audio correspondence,
AudioScape-TTA represents realistic soundscapes through structured audio
components and enables fine-grained semantic evaluation through
requirement-level verification.

As illustrated in Figure~\ref{fig:overview}, AudioScape-TTA consists of two main components:
(1) a structured soundscape benchmark that organizes audio scenes into explicit audio components and characterizes sample complexity through event density and structural complexity; and
(2) an audio-grounded rubrics-based evaluation system that converts structured soundscape descriptions into fine-grained semantic rubrics for assessing generated audio.
Together, these components enable interpretable analysis of TTA systems beyond global text--audio similarity metrics.

\subsection{Structured Soundscape Benchmark}

Existing TTA benchmarks mainly provide caption-level descriptions, which are insufficient for analyzing realistic soundscapes containing multiple interacting audio components. To address this limitation, AudioScape-TTA introduces a structured representation that organizes audio scenes according to their semantic composition, including environmental context, background music, sound effects, and speech.

\noindent\textbf{Data Collection and Annotation.}
AudioScape-TTA is constructed from real-world audio scenes extracted from movie and television productions. We first segment the original audio tracks into short clips and perform keyword-based retrieval to identify candidate samples covering diverse acoustic events and soundscape compositions.

For each candidate clip, we adopt a semi-automatic annotation pipeline in which an LLM is used to assist the initial drafting of structured descriptions and component annotations. All annotations are subsequently reviewed and refined by trained annotators, who verify semantic correctness, revise inaccurate descriptions, refine annotations for scene context, sound effects, background music, and speech, and remove noisy or ambiguous samples to ensure annotation quality and consistency.
The final benchmark is obtained through a three-stage pipeline consisting of candidate retrieval, LLM-assisted annotation, and human quality assurance, where human verification serves as the final authority for all benchmark annotations.

\noindent\textbf{Structured Soundscape Representation.}
Each sample in AudioScape-TTA is represented as a structured soundscape
description consisting of a scene context and modality-specific audio components:
1)\textbf{Scene}: the overall acoustic environment or contextual setting of the audio clip;
2) \textbf{Background Music (BGM)}: musical elements such as instruments, style, and mood;
3) \textbf{Sound Effects (SFX)}: individual acoustic events and environmental actions;
4) \textbf{Speech}: spoken content and speaker-related information.

The scene component provides high-level contextual information for describing realistic audio environments, while SFX, BGM, and speech define the explicit audio components evaluated in our semantic verification framework. This representation captures the compositional nature of realistic soundscapes and enables analysis beyond isolated event generation.
Here, composition refers to the clip-level organization and co-occurrence of predefined audio components. This representation focuses on semantic composition among soundscape components rather than explicitly modeling fine-grained temporal ordering or causal interactions among individual events.
\noindent\textbf{Complexity-aware Characterization.}
To enable complexity-aware evaluation, AudioScape-TTA characterizes each sample using two complementary dimensions: \emph{event density} and \emph{structural complexity}. Event density measures the amount of acoustic content in a clip, whereas structural complexity characterizes the diversity and clip-level composition of active audio categories. Together, these two dimensions capture complementary aspects of soundscape complexity, enabling systematic analysis of model robustness from simple scenes to dense and compositionally complex soundscapes.
Based on the normalized combination of these two dimensions, we partition samples into three complexity levels: \textit{Easy}, \textit{Medium}, and \textit{Hard}. Figure~\ref{fig:overview} summarizes the distributions of event density and structural complexity, while the complete definitions of the two measures and the complexity partition strategy are provided in supplementary material.

\noindent\textbf{Data Statistics.}
As summarized in Table~\ref{tab:statis}, AudioScape-TTA contains
2,258 audio-text instances with 25,707 fine-grained semantic rubrics.
Each audio has an average duration of 9.95 seconds, and each textual
annotation contains 32.45 words on average. Each sample includes
approximately 4 annotated semantic events, enabling fine-grained evaluation
of event realization, attribute fidelity, and speech-related requirements.
In particular, speech annotations are provided for 358 clips, including
target-utterance evaluation and speech-attribute evaluation covering
speaker and linguistic characteristics. Detailed statistics of modality
distribution, event categories, and annotation characteristics are provided
in supplementary material.

\begin{table*}[t]
\centering
\footnotesize
\setlength{\tabcolsep}{4.5pt}
\begin{tabular}{l|c|c|cc|ccccc|c}
\toprule
\multirow{2}{*}{\textbf{Model}}
& \multirow{2}{*}{\textbf{Year}}
& \multirow{2}{*}{\textbf{Params}}
& \multicolumn{2}{c|}{\textbf{Overall}}
& \multicolumn{2}{c}{\textbf{Event Presence}}
& \multicolumn{2}{c}{\textbf{Event Attribute}}
& \textbf{Speech Content}
& \textbf{Text-Audio} \\
\cmidrule(lr){4-5}
\cmidrule(lr){6-7}
\cmidrule(lr){8-9}
\cmidrule(lr){10-10}
\cmidrule(lr){11-11}
&
&
& \textbf{SR (\%)}$\uparrow$
& \textbf{Conf.}
& \textbf{SR (\%)}$\uparrow$
& \textbf{Conf.}
& \textbf{SR (\%)}$\uparrow$
& \textbf{Conf.}
& \textbf{SCCA@0.60}$\uparrow$
& $\mathbf{CLAP}_{\mathrm{MS}}\uparrow$ \\
\midrule

AudioLDM 2
& 2023 & 1.5B
& 66.92 & 93.85
& 68.43 & 94.05
& 66.57 & 93.71
& 0.00
& 0.4193 \\

Make-An-Audio 2
& 2023 & 937M
& 61.33 & 92.99
& 61.48 & 93.26
& 61.89 & 92.80
& 0.00
& 0.3321 \\

AudioGen
& 2023 & 1.5B
& 66.53 & 93.62
& 68.40 & 93.99
& 65.92 & 93.35
& 0.00
& 0.3664 \\

Tango 2
& 2024 & 866M
& 76.82 & 93.33
& \underline{80.84} & 93.97
& 74.78 & 92.88
& 0.00
& 0.4145 \\

TangoFlux
& 2024 & 515M
& 76.77 & 94.27
& 80.55 & 94.56
& 74.89 & 94.07
& 0.00
& \underline{0.4736} \\

EzAudio
& 2025 & 875M
& \underline{78.20} & 94.14
& 80.83 & 94.38
& \underline{77.17} & 93.97
& 0.00
& 0.4713 \\

MAGNeT
& 2024 & 1.5B
& 53.39 & 92.44
& 49.89 & 92.38
& 56.50 & 92.48
& 0.00
& 0.2827 \\

Stable Audio Open
& 2025 & 1.1B
& 65.95 & 93.59
& 67.57 & 93.98
& 65.51 & 93.32
& 0.00
& 0.3480 \\

MMAudio
& 2025 & 1.03B
& 64.73 & 94.15
& 67.80 & 94.52
& 63.24 & 93.89
& 0.00
& \textbf{0.5169} \\

Foley-Omni
& 2026 & 5.5B
& \textbf{79.62} & 94.05
& \textbf{81.74} & 94.66
& \textbf{78.34} & 93.61
& \underline{57.06}
& 0.4378 \\

Omni2Sound
& 2026 & 1.33B
& 77.59 & 94.47
& 79.98 & 94.86
& 76.72 & 94.20
& 0.00
& 0.4370 \\

AudioStory
& 2026 & 4.36B
& 69.03 & 93.04
& 70.09 & 93.36
& 69.02 & 92.80
& 0.00
& 0.3989 \\

Dasheng AudioGen
& 2026 & 2.19B
& 74.50 & 94.17
& 74.86 & 94.27
& 74.21 & 94.09
& \textbf{77.30}
& 0.4139 \\

\bottomrule
\end{tabular}

\caption{
Fine-grained semantic evaluation results on AudioScape-TTA.
Best and second-best results are highlighted in \textbf{bold} and
\underline{underline}, respectively.
}
\label{tab:overall}
\end{table*}


\begin{table}[htbp]
    \centering
    \small
    \begin{tabular}{lc}
    \toprule
        \textbf{Statistics} & \textbf{Number} \\
    \midrule
      Total Samples & 2,258 \\
      Total Semantic Rubrics & 25,707 \\
      Avg. Audio Duration & 9.95 sec \\
      Avg. Annotation Length & 32.45 words \\
      Avg. Semantic Rubrics & 11.39 \\
      Avg. Events & 4 \\
    \midrule
      \multicolumn{2}{c}{\textbf{Speech-related Statistics}} \\
    \midrule
      Speech-annotated Clips & 358 (15.85\%) \\
      Clips with Target Utterances & 83 \\
      Speech-content Rubrics & 163 \\
      Speech-attribute Rubrics & 931 \\
    \bottomrule 
    \end{tabular}
    \caption{
    Key statistics of the AudioScape-TTA benchmark.
    }
    \label{tab:statis}
\end{table}
\subsection{Rubric-based Audio-grounded Evaluation}

Given a structured soundscape description, the goal of evaluation is to determine whether each predefined semantic requirement is satisfied by the generated audio. To this end, we propose a rubric-based audio-grounded evaluation framework that converts each structured annotation into a fixed set of semantic verification rubrics and evaluates them through audio-language understanding. As illustrated in Figure~\ref{fig:overview}, the framework consists of three stages: (1) structured rubric construction, (2) audio-grounded verification, and (3) hierarchical score aggregation.

\noindent\textbf{Structured Rubric Construction.}
Given a structured description, we employ Qwen3.5-27B~\cite{qwen3.5} to extract a hierarchical semantic representation:
\[
\text{Sample}
\rightarrow
\text{Modality}
\rightarrow
\text{Event}
\rightarrow
\text{Attribute}.
\]
Based on this hierarchy, we construct a fixed set of binary semantic rubrics during benchmark construction. Each rubric defines a fixed semantic verification criterion and is shared across all evaluated models, ensuring consistent and auditable comparison:
(1) \textbf{Event Presence}, which verifies whether a required sound event is generated;
(2) \textbf{Event Attribute}, which verifies whether the generated event satisfies the specified acoustic properties;
(3) \textbf{Speech Content}, which verifies whether synthesized speech preserves the intended linguistic content.

\noindent\textbf{Audio-grounded Verification.}
Each semantic rubric is evaluated using a verifier according to its semantic type. Event-presence and event-attribute rubrics are verified by Qwen3-Omni-Instruct~\cite{xu2025qwen3}, while speech-content rubrics are evaluated by Qwen3-ASR~\cite{shi2026qwen3} through transcript matching with the annotated target utterance.

\noindent\textbf{Hierarchical Score Aggregation.}
Each fixed semantic rubric is independently verified and contributes equally to the final benchmark score, which is computed as the micro-average satisfaction rate over all rubrics.
 Beyond the overall score, AudioScape-TTA provides hierarchical evaluation from three complementary perspectives. \textbf{Rubrics-based evaluation} measures semantic instruction following across event presence, event attributes, and speech content. \textbf{Modality-level evaluation} analyzes performance on sound effects, background music, and speech. \textbf{Complexity-aware evaluation} further assesses model robustness across Easy, Medium, and Hard soundscapes characterized by different levels of event density and structural complexity. Together, these analyses provide interpretable diagnosis of semantic instruction following beyond a single aggregate score. Detailed metric definitions are provided in supplementary material.

\section{Experiments}

\subsection{Experimental Setup}

\noindent\textbf{Evaluated Models.}
We evaluate 13 open-source TTA models on AudioScape-TTA, covering diverse generation paradigms. The evaluated models include AudioLDM2~\cite{liu2023audioldm}, Make-An-Audio 2~\cite{huang2023make2}, MMAudio~\cite{cheng2025mmaudio}, Tango 2~\cite{majumder2024tango}, TangoFlux~\cite{hung2024tangoflux}, EzAudio~\cite{hai2025ezaudio}, MAGNeT~\cite{ziv2024masked}, AudioGen~\cite{kreuk2022audiogen}, Stable Audio Open~\cite{evans2025stable}, Foley-Omni~\cite{tao2026foley}, Omni2Sound~\cite{dai2026omni2sound}, AudioStory~\cite{guo2026audiostory}, and Dasheng AudioGen~\cite{mei2026dasheng}. 
Detailed model architectures and generation paradigms are summarized in Table~\ref{tab:model_architectures} in the supplementary material.

\noindent\textbf{Evaluation Protocol.}
We evaluate generated audio from two complementary perspectives:
fine-grained requirement satisfaction and global generation quality.
For fine-grained evaluation, we report requirement satisfaction rates
over overall, modality-specific, complexity-level, and QA-type subsets.
Event-presence and event-attribute requirements are evaluated by
Qwen3-Omni-Instruct~\cite{xu2025qwen3} using binary verification QA,
where each rubric checks whether a specified target requirement is realized
in the generated audio. Speech-content requirements are evaluated by
Qwen3-ASR~\cite{shi2026qwen3} based on transcript matching.

Since all semantic rubrics in the benchmark are formulated as positive
requirements, the resulting score measures the proportion of requested
requirements successfully satisfied by the generated audio rather than
general semantic satisfaction rate or hallucination-free generation.
Accordingly, we denote a metric as Satisfaction Rate (SR).
For speech-content requirements, we introduce Speech Content Coverage
Satisfaction Rate at threshold 0.60 (SCCA@0.60) to convert transcript matching into binary requirement satisfaction. 
Word Error Rate (WER) and Character Error Rate (CER) are further
reported in the supplementary material as complementary transcription fidelity diagnostics,
but are not used for the primary semantic evaluation because many TTA systems
do not explicitly generate speech content, resulting in degenerate ASR-based
scores.

For global audio quality assessment, we include Audiobox-Aesthetic
metrics, including Content Enjoyment (CE),  Content Usefulness (CU),
Production Complexity (PC), and Production Quality (PQ)~\cite{tjandra2025meta};
Text-audio alignment score
($\mathbf{CLAP}_{\mathrm{MS}}$)~\cite{elizalde2024natural};
and distribution-based metrics including Fréchet Audio Distance (FAD),
Kullback--Leibler (KL) divergence, and Inception Score (ISC).
Detailed metric definitions and implementation details are provided in
supplementary material.

\begin{figure}[t]
    \centering
    \includegraphics[width=1\linewidth]{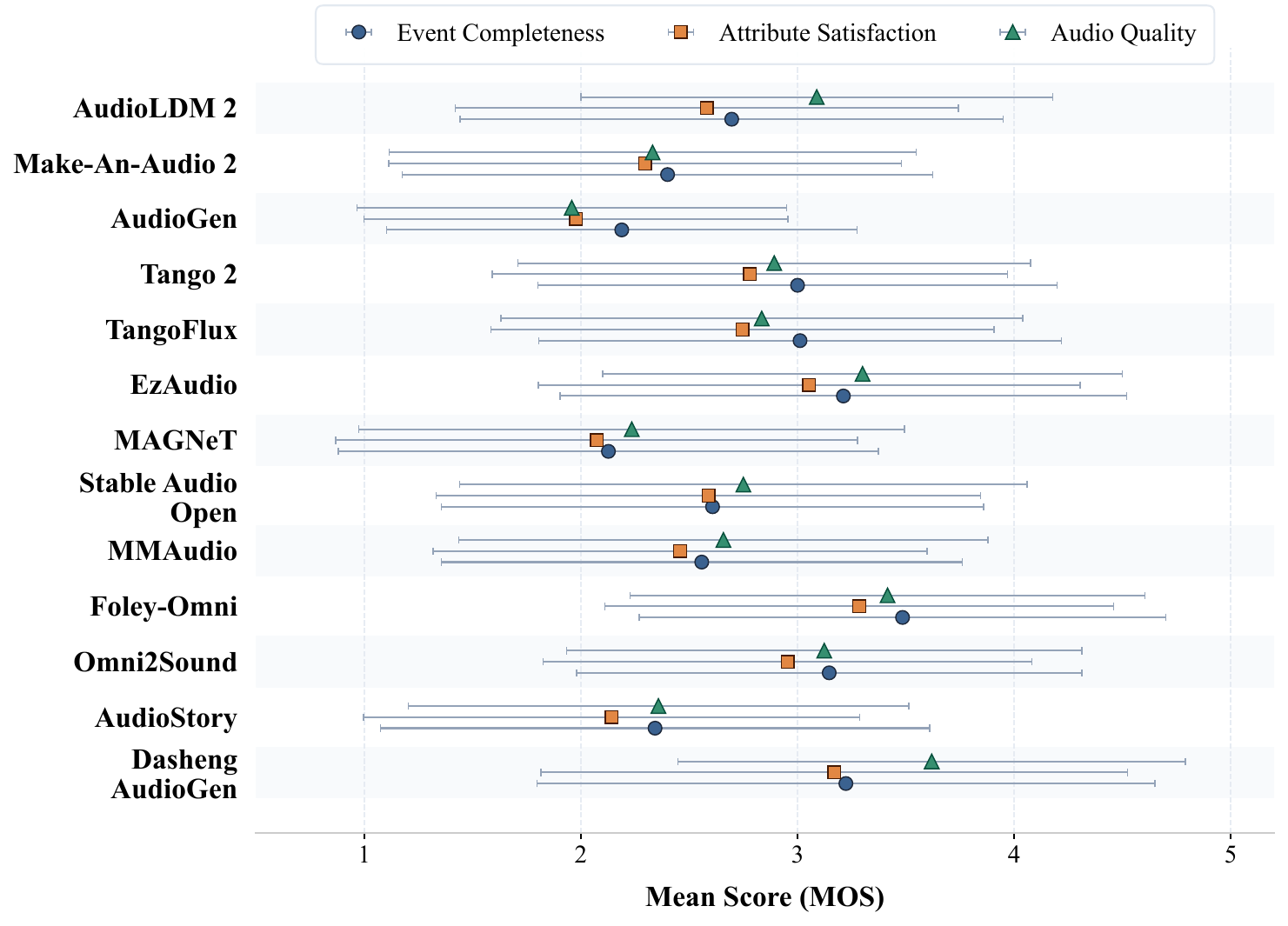}
    \caption{User study results on the AudioScape-TTA benchmark. Markers indicate mean human ratings for event completeness, attribute satisfaction rate, and audio quality, while error bars denote standard deviations.}
    \label{fig:user_study}
\end{figure}

\subsection{Fine-grained Semantic Evaluation}

\noindent\textbf{Overall Results.}
Table~\ref{tab:overall} summarizes the fine-grained semantic evaluation results.
Among all evaluated models, Foley-Omni achieves the best overall performance,
obtaining 79.62\% SR and ranking first on both event-presence
(81.74\%) and event-attribute (78.34\%) evaluation.
Two observations are noteworthy. First, event-attribute satisfaction rate consistently
lags behind event-presence satisfaction rate, indicating that accurately generating a
target event is substantially easier than faithfully controlling its acoustic
attributes. Second, speech-content generation remains a major challenge.
Except for Foley-Omni and Dasheng AudioGen, all evaluated models achieve
0\% SCCA, revealing that current open-source TTA systems largely fail to
preserve linguistic content despite producing speech-like audio.

\noindent\textbf{Modality-specific Analysis.}
Figure~\ref{fig:modality_difficulty} (a) reveals clear modality-specific
differences among current TTA systems. Foley-Omni achieves the strongest
performance on sound effects, while EzAudio performs best on music. Speech is
considerably more challenging than the other modalities, with Dasheng AudioGen
being the only model achieving competitive speech-content satisfaction rate.
Performance further degrades on multi-modal compositions that jointly involve
sound effects, music, and speech (Table~\ref{tab:modality_composition} in the
supplementary material), suggesting that modeling interactions among audio
components remains a key challenge for current TTA systems. Foley-Omni and
Dasheng AudioGen show the strongest robustness under these compositional
settings.

\noindent\textbf{Complexity-aware Analysis.}
Figure~\ref{fig:modality_difficulty} (b) shows that the semantic satisfaction rate consistently
degrades as complexity increases. While several models perform
competitively on easy subsets, their performance drops substantially on medium
and hard subsets. Foley-Omni exhibits strong robustness, particularly on
the hard subset, indicating superior semantic grounding under dense
compositional soundscapes.


\subsection{Comparison with Conventional Evaluation Metrics}

Table~\ref{tab:overall} additionally reports $\mathbf{CLAP}_{\mathrm{MS}}$ as
a conventional global text--audio alignment metric for comparison. The ranking
under CLAP-based similarity is not consistent with fine-grained
rubrics-based semantic satisfaction. For example, MMAudio achieves the highest
$\mathbf{CLAP}_{\mathrm{MS}}$ score (0.5169) but obtains only 64.73\% semantic
requirement satisfaction rate, indicating that strong global text--audio
alignment does not necessarily guarantee faithful realization of individual
semantic requirements. In contrast, Foley-Omni achieves the highest semantic
satisfaction rate while ranking lower under CLAP-based alignment.
These results suggest that global embedding-based metrics and
rubrics-based evaluation capture complementary aspects of TTA generation:
CLAP measures overall text--audio correspondence, whereas rubrics-based
evaluation directly verifies whether individual prompt-specified requirements
are satisfied.

We further report perceptual quality and distribution-based metrics, including
Audiobox-Aesthetic scores, FAD, KL divergence, and ISC, in the supplementary
material.

\subsection{User Study}

To obtain human judgments for validating metric reliability, we conduct a user
study on AudioScape-TTA. We randomly select 10 prompts covering diverse
modality compositions and complexity levels, and collect generated audio from
all 13 evaluated models, resulting in 130 samples. Each sample is rated by 26
participants on three dimensions: \textit{event completeness},
\textit{attribute satisfaction rate}, and \textit{audio quality}. Event completeness
measures whether required events are generated, attribute satisfaction rate evaluates
whether fine-grained acoustic properties are preserved, and audio quality
assesses perceptual naturalness and generation artifacts. Participants rate each dimension using a 5-point Likert scale.

As shown in Figure~\ref{fig:user_study}, human judgments closely agree with
the fine-grained QA evaluation. Foley-Omni achieves the highest scores in event
completeness and attribute satisfaction rate, consistent with its leading semantic
following performance. Dasheng AudioGen obtains the highest audio quality
score, indicating stronger perceptual quality among the evaluated models.

\subsection{Human Alignment of Evaluation Metrics}

We quantify the agreement between human judgments and automatic metrics using model-level Spearman correlation across the 13 evaluated TTA models. Human evaluation provides ratings for event completeness, attribute satisfaction, and audio quality. To summarize semantic instruction following into a single target, we define a \emph{composite human semantic score} (Semantic$^\dagger$) as the equal-weight average of event completeness and attribute satisfaction. Audio quality is excluded because it measures perceptual preference rather than semantic correctness.

Table~\ref{tab:human_auto_corr} shows that the our rubric-based metrics achieve strong alignment with corresponding human judgments. Event SR exhibits a strong correlation with human event-completeness ratings, while Attribute SR closely matches human attribute-satisfaction ratings. The aggregated Overall SR achieves the highest correlation with the composite human semantic score, substantially outperforming $\mathbf{CLAP}_{\mathrm{MS}}$, which measures global text--audio similarity. These results demonstrate that rubric-based evaluation better captures human-perceived semantic instruction following than global embedding-based similarity metrics in our evaluation setting.

\begin{table}[t]
\centering
\small
\setlength{\tabcolsep}{5pt}
\begin{tabular}{lccc}
\toprule
\textbf{Human Judgment} & \textbf{Metric} & $\rho$ & 
\textbf{$p_{\mathrm{Holm}}$} \\
\midrule

Completeness & Event SR & 0.743 & 0.0072 \\

Attribute & Attribute SR & 0.825 & 0.0003 \\

\multirow{2}{*}{Semantic$^\dagger$} 
& Overall SR & \textbf{0.879} & 0.0015 \\

& $\mathbf{CLAP}_{\mathrm{MS}}$ & 0.312 & 0.0739 \\

\bottomrule
\end{tabular}

\caption{
Model-level Spearman correlations between automatic evaluation metrics and human judgments across 13 TTA models using the same samples as the user study.
All two-sided $p$ values are Holm-adjusted over the four correlation tests.
Bold indicates the highest correlation coefficient.
Semantic$^\dagger$ denotes the equal-weight average of event completeness and attribute satisfaction.
}
\label{tab:human_auto_corr}
\end{table}

\subsection{Discussion}

Our analysis reveals several key challenges in current text-to-audio
generation. 
First, semantic event recognition and fine-grained acoustic
control remain fundamentally different capabilities. While current models can
often generate the target sound events, reproducing the associated attributes
and detailed acoustic characteristics remains challenging, suggesting that
future TTA models require stronger compositional and attribute-aware
generation mechanisms.
Second, speech generation remains a distinct challenge compared with other
audio modalities. Preserving linguistic content, speaker identity, and
acoustic realism simultaneously requires modeling multiple aspects of speech,
which is not yet reliably achieved by unified TTA systems.
Third, current models struggle with highly compositional soundscapes involving
multiple events and interacting modalities. The degradation under increasing
complexity suggests that existing systems are still primarily optimized for
individual sound generation rather than structured audio scene composition.
Finally, our human correlation analysis highlights the importance of
fine-grained evaluation for future TTA research. Global metrics remain useful
for assessing perceptual quality and coarse similarity, but they provide
limited insight into specific semantic failures. Structured evaluation enables
more interpretable diagnosis of model capabilities and facilitates future
progress toward controllable audio generation.

\section{Conclusion}

In this work, we introduced \textbf{AudioScape-TTA}, a structured and
complexity-aware benchmark and evaluation framework for fine-grained
text-to-audio generation. By representing soundscapes through
modality-aware semantic structures and evaluating generation results through
rubric-based verification, AudioScape-TTA enables interpretable
analysis beyond conventional global similarity metrics. Extensive experiments
on 13 open-source TTA models reveal persistent limitations in fine-grained
attribute realization, speech-content preservation, and compositional
soundscape generation. Human validation further demonstrates that our
evaluation framework aligns well with human semantic judgments, highlighting
the importance of requirement-level evaluation for assessing TTA instruction
following.
We hope AudioScape-TTA can facilitate the development of more controllable and
semantically faithful audio generation systems. Future extensions may explore
richer soundscape representations with temporal, spatial, and causal relations,
as well as more advanced evaluation protocols for complex audio interactions
and long-form generation scenarios.

\bibliography{aaai2027}

\clearpage


\section{Additional Dataset Details}

\subsection{Annotation Pipeline}

The construction of AudioScape-TTA follows a three-stage annotation pipeline.

First, candidate audio clips are retrieved using keyword-based filtering from a large-scale video repository. Second, Gemini-2.5-Pro is used to generate initial descriptions and
candidate structured annotations, including modality-level events and
attributes. Third, human annotators verify and refine these annotations,
correct semantic errors, and remove ambiguous samples.

This procedure ensures both annotation scalability and semantic reliability.

\subsection{Complexity Score Definition}
\label{app:complexity}

We characterize sample complexity using two complementary dimensions: \emph{event density} and \emph{structural complexity}. While event density measures the amount of acoustic content contained in a clip, structural complexity characterizes the diversity and clip-level composition of active audio categories. Together, they capture two complementary aspects of generation complexity.

\paragraph{Event Density.}

Event density measures how much acoustic content is contained in a soundscape. It is defined as

\begin{equation}
\text{Density}
=
N_{\mathrm{SFX}}
+
N_{\mathrm{Speech}}
+
N_{\mathrm{BGM}},
\end{equation}

where $N_{\mathrm{SFX}}$ denotes the number of annotated SFX event instances in a clip, with repeated occurrences counted separately. In contrast, $N_{\mathrm{Speech}} \in \{0,1\}$ and $N_{\mathrm{BGM}} \in \{0,1\}$ are binary indicators denoting the presence of speech and background music, respectively.

This asymmetric definition reflects the different temporal characteristics of the three audio categories. SFX typically consists of discrete and countable event instances, whereas speech usually forms a continuous dialogue layer and BGM serves as a sustained background layer. Counting individual speech utterances or music segments would depend heavily on annotation granularity and could artificially inflate the density score. We therefore count SFX at the event-instance level while representing speech and BGM solely by their clip-level presence. Note that density does not represent physical loudness or duration; it only measures the number of annotated semantic components.

\begin{table*}[htbp]
\centering
\small
\setlength{\tabcolsep}{5pt}
\begin{tabular}{cp{7.2cm}p{6.5cm}}
\toprule
\textbf{Complexity} & \textbf{Annotation}& \textbf{Semantic Rubrics} \\
\midrule

Easy &
\textit{[Scene] Outdoor water source environment; 
[SFX] Gentle continuous gurgling and trickling of flowing water; 
[SFX] Water pouring out of the spout.}& \textbf{Event presence: } \textit{1. Is there an outdoor water source environment?
 2. Is there flowing water? 3. Is there water pouring out of the spout?} 
 
 \textbf{Event attributes:} \textit{1. Is the sound of flowing water characterized by gurgling and trickling?
2. Is the sound of flowing water gentle and continuous?}\\
\hline

Medium &
\textit{[Scene] Indoor cafe/kitchen environment; 
[BGM] Slow gentle sentimental piano music; 
[SFX] Grinding/blending machinery sound; 
[SFX] Subtle clinks and food preparation sounds; 
[SFX] Bowl being placed on counter.} &  \textbf{Event presence: } \textit{1. Is there an indoor cafe/kitchen environment present? 2. Is there a grinding or blending machinery sound present? 3.  Are there subtle clinks and food-preparation sounds? 4. Is there a bowl being placed on counter present? 5. Is there piano music present?} 

 \textbf{Event attributes:} \textit{1. Is the sound of food preparation characterized by subtle clinks? 2. Is the mood of the piano music slow, gentle, and sentimental? 3. Is the instrument of the slow gentle sentimental music a piano?}\\
\hline
Hard &
\textit{[Scene] Indoor closed environment; 
[BGM] Tense atmospheric dramatic score with ominous synthesized tones; 
[SFX] Male speaking; Thunder and lightning; Baby crying; 
[Speaker A, elderly male, hoarse voice]: ``Tell the story. He said he could hear that baby cry just as plain...''} & \textbf{Event presence: } \textit{1. Is there an indoor closed environment?
2. Is there a male speaking?
3. Is there thunder and lightning?
4. Is there a baby crying?
5. Are ominous synthesized tones present?
} 

\textbf{Speech content:} \textit{Does the speech include the phrase 'tell the story. He said he could hear that baby cry just as plain. And um...?}
 
 \textbf{Event attributes:} \textit{1. Is the speaker male?
2. Is the speaker elderly?
3. Is the speaker speaking in  hoarse voice?
4. Is the source of the crying referred to as a baby?
5. Is the action or content of the crying described as crying?
6. Is the genre of the score described as atmospheric dramatic?
7. Is the mood of the score described as tense?
8. Is the instrument of the score described as synthesized tones?
}\\

\bottomrule
\end{tabular}
\caption{Examples of structured annotation and semantic rubrics across different complexity levels.}
\label{tab:difficulty_examples}
\end{table*}

\begin{figure*}[htbp]
    \centering
    \includegraphics[width=0.95\linewidth]{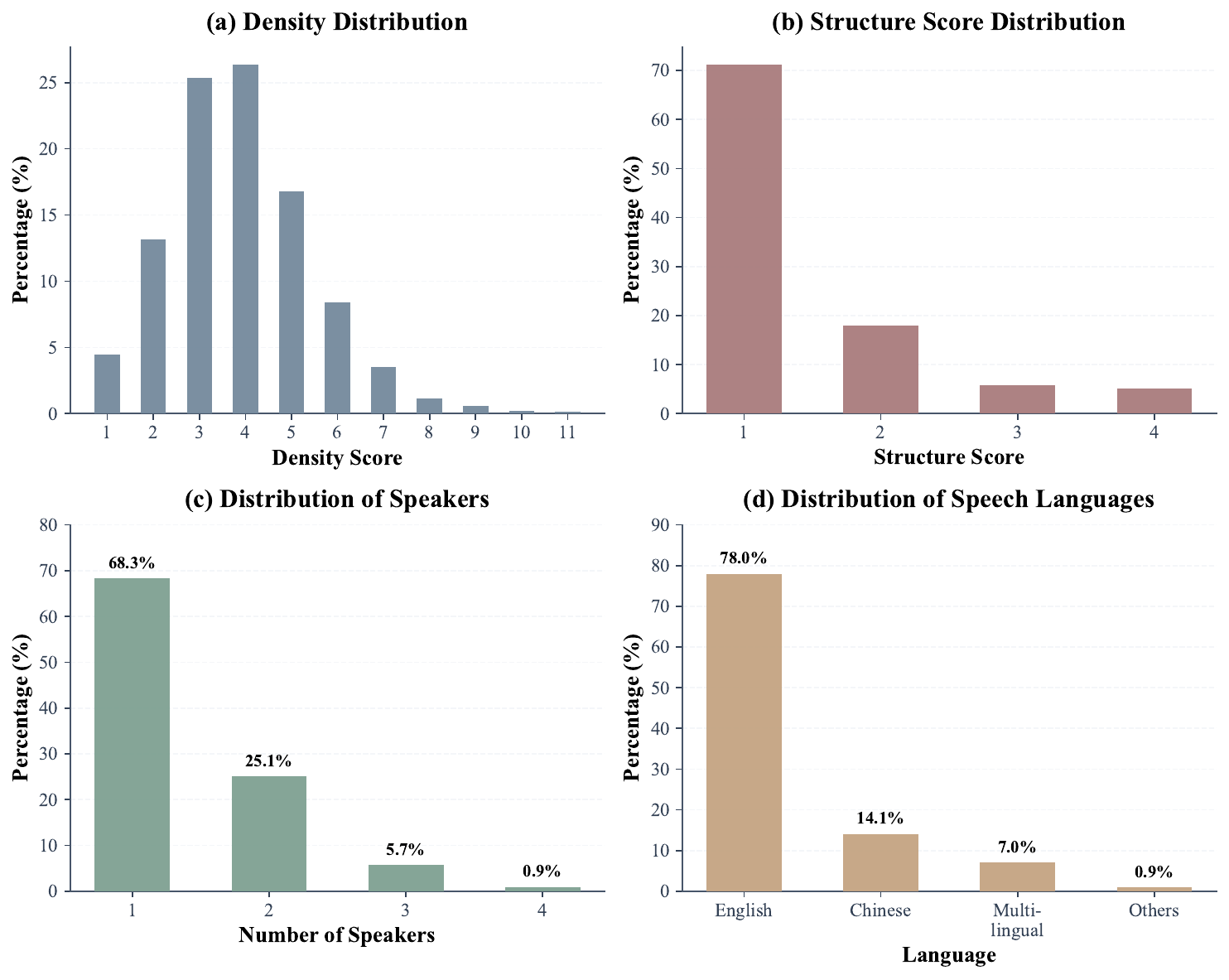}
    \caption{\textbf{Statistics of the proposed AudioScape-TTA benchmark.} 
(a) Density distribution reflects the combination of discrete SFX-instance
counts and clip-level presence indicators for speech and BGM.
(b) Structure score distribution characterizes the complexity and organization of multi-event soundscapes. 
(c) and (d) summarize speaker number and speech language distributions, highlighting the diversity of speech-related scenarios. 
These statistics demonstrate the multimodal diversity and compositional complexity of the benchmark.}
    \label{fig:statics_appendix}
\end{figure*}

\paragraph{Structural Complexity.}

Structural complexity measures how diverse the composition of a soundscape is, independent of the number of event instances. It is defined as

\begin{equation}
\text{Structure\_Score}
=
S_{\mathrm{modality}}
+
S_{\mathrm{cooccur}},
\end{equation}

where $S_{\mathrm{modality}}$ denotes the number of active audio categories among 
\textit{Speech}, \textit{SFX}, and \textit{BGM}. We treat speech as a distinct 
modality because, unlike non-speech audio components, speech generation requires 
not only acoustic realization but also preservation of linguistic content and 
speaker-related characteristics.

The clip-level co-occurrence term $S_{\mathrm{cooccur}} \in \{0,1\}$ is defined as

\begin{equation}
S_{\mathrm{cooccur}}
=
\mathbb{I}
\left[
\mathrm{Speech}
\land
(\mathrm{SFX} \lor \mathrm{BGM})
\right],
\end{equation}

where $\mathbb{I}[\cdot]$ is the indicator function. It equals $1$ when speech 
co-occurs with at least one non-speech audio category (SFX or BGM), and $0$ 
otherwise. Therefore, Structure\_Score ranges from 0 to 4, where larger values 
indicate more diverse modality combinations.

Unlike event density, structural complexity does not depend on the number of SFX 
instances. Instead, it captures the diversity of active audio categories and 
whether heterogeneous audio components must be generated jointly. We explicitly 
consider speech-related co-occurrence because unified speech-aware audio 
generation introduces additional requirements beyond non-speech sound synthesis, 
including linguistic content preservation, speaker identity modeling, and 
coordination between vocal and non-vocal components. Therefore, soundscapes 
containing speech together with SFX or BGM typically require joint modeling of 
semantic, acoustic, and linguistic factors.

The proposed structural complexity is designed as a practical benchmark-oriented 
complexity heuristic rather than a claim that speech is universally more complex 
than other audio categories. The co-occurrence term characterizes category-level 
composition only; it does not explicitly model temporal overlap, event ordering, 
causal relations, or semantic interactions among components.
Accordingly, the proposed complexity characterization captures clip-level
audio-category composition rather than explicit temporal overlap, event
ordering, causal relations, or semantic interactions among individual events.
References to compositional difficulty in this paper should therefore be
interpreted under this benchmark-specific Density--Structure characterization.

\paragraph{Complexity Score.}

Since event density and structural complexity are measured on different scales, we first normalize them using z-score normalization and then compute the overall complexity score as

\begin{equation}
C
=
Z(\text{Density})
+
Z(\text{Structure\_Score}),
\end{equation}

where $Z(\cdot)$ denotes z-score normalization.

Finally, samples are partitioned into three complexity levels according to

\[
\begin{cases}
\text{Easy}, & C\leq -1,\\
\text{Medium}, & -1<C<1,\\
\text{Hard}, & C\geq 1.
\end{cases}
\]
The thresholds are selected based on standardized complexity scores to
separate low-, medium-, and high-complexity samples while maintaining
sufficient samples in each subset.
This formulation enables complexity-aware evaluation by jointly considering both the amount of acoustic content and the compositional diversity of structured soundscapes.

\paragraph{Complementarity Analysis.}
Across all 2,258 samples, event density and structural complexity exhibit only weak correlation (Pearson $r = 0.221$; Spearman $\rho = 0.153$), indicating that the two metrics capture complementary aspects of audio-scene complexity. Specifically, event density measures the amount of acoustic content, whereas structural complexity characterizes the diversity and co-occurrence patterns of audio components.

\begin{table*}[t]
\centering
\small
\setlength{\tabcolsep}{5pt}
\begin{tabular}{lrrrrrr}
\toprule
\textbf{Subset} &
\textbf{\# Samples} &
\textbf{Avg.\ SFX} &
\textbf{Speech (\%)} &
\textbf{BGM (\%)} &
\textbf{Multi-comp.\ (\%)} &
\textbf{Avg.\ Rubrics} \\
\midrule
Easy   & 732   & 2.36 & 0.00  & 10.52 & 0.00  & 7.62 \\
Medium & 1,103 & 3.91 & 0.73  & 25.84 & 26.38 & 11.92 \\
Hard   & 423   & 4.48 & 56.03 & 56.50 & 85.11 & 16.51 \\
\bottomrule
\end{tabular}
\caption{
Composition statistics of the Easy, Medium, and Hard subsets.
Avg.\ SFX denotes the mean number of annotated SFX instances per clip.
Speech and BGM indicate the percentages of clips containing the corresponding
modality. Multi-comp.\ denotes the percentage of clips containing at least two
modalities among Speech, SFX, and BGM. Avg.\ Rubrics is the average number of
QA rubrics per clip in the final released benchmark.
}
\label{tab:difficulty_composition}
\end{table*}

\subsection{Data Statistics}
\label{app:statistics}
To further demonstrate the diversity and complexity of the proposed AudioScape-TTA benchmark, we provide additional annotation examples and dataset statistics in the supplementary material. 
Table~\ref{tab:difficulty_examples} presents representative samples with different complexity levels, where each audio scene is described through a hierarchical annotation scheme covering scene context, sound effects (SFX), background music (BGM), speaker attributes, and speech content. 
The corresponding semantic rubrics further decompose these complex descriptions into fine-grained evaluation criteria, enabling detailed assessment of event presence, attributes, and speech semantics. 
As the complexity increases, samples contain richer modality combinations, more discrete SFX instances, and more complex clip-level audio-category composition, ranging from simple isolated sound sources to challenging scenarios involving simultaneous environmental sounds, music, and speech.

Moreover, Figure~\ref{fig:statics_appendix} summarizes the benchmark statistics from multiple perspectives. 
The density and structure score distributions characterize the complexity of multi-event soundscapes, reflecting the number of co-occurring audio elements and their structural organization. 
The speaker and language distributions further demonstrate the diversity of speech-related scenarios. 
Together, these results highlight that AudioScape-TTA covers diverse audio modalities and challenging compositional sound environments, providing a comprehensive testbed for evaluating TTA generation models.

 Table~\ref{tab:difficulty_composition} summarizes the composition of the Easy, Medium, and Hard subsets. As the difficulty level increases, audio scenes become progressively more complex, with more annotated SFX events, higher proportions of speech and background music, more multimodal compositions, and a larger number of QA rubrics per clip. These statistics confirm that the proposed difficulty partition reflects increasing compositional complexity rather than merely differences in sample count.

\subsection{Semantic Rubrics Construction Details}

We construct rubrics pairs from structured event-level annotations using a deterministic target-generation pipeline followed by natural-language rewriting. The semantic target of each rubrics pair is derived directly from the annotation, while the LLM is only used to rewrite the target into a concise listener-answerable yes/no question. Therefore, the LLM does not decide the answer or introduce new semantic targets. All rubrics pairs use the same two-choice format, \texttt{A: Yes} and \texttt{B: No}, and the reference answer is set to \texttt{A} because each question is generated from a positive annotated event or attribute.

\paragraph{Rubrics Types.}
We generate three types of rubrics pairs:
\begin{itemize}
    \item \textbf{Event presence}: verifies whether a non-speech sound event, background music, or acoustic scene is present in the audio.
    \item \textbf{Event attribute}: verifies whether a specific event-level acoustic attribute is realized, such as sound source, material, intensity, temporal pattern, spatial property, instrument, emotion, speaker identity, or language.
    \item \textbf{Speech content}: verifies whether the spoken content of a speech event is preserved.
\end{itemize}

\paragraph{Structured rubrics schema.}
Each QA rubrics pair stores both the natural-language question and its structured evidence:
\begin{lstlisting}
{
  "qa_id": string,
  "qa_type": "event_presence | event_attribute | speech_content",
  "answer_type": "two_choice",
  "question": string,
  "choices": {"A": "Yes", "B": "No"},
  "answer": "A",
  "modality": "SFX | BGM | Speech | Scene",
  "event_id": int,
  "event_text": string,
  "attribute_field": string,
  "attribute_value": string,
  "evidence": object
}
\end{lstlisting}

\paragraph{Target generation.}
For each annotated event, we first construct structured QA targets. For non-speech events, we create an \texttt{event\_presence} target. For all events, we select informative attributes according to modality-specific priority lists and create \texttt{event\_attribute} targets. Generic or uninformative values are removed. For speech events, if the \texttt{action\_or\_content} field contains explicit spoken content, we create a \texttt{speech\_content} target.

The attribute priorities are:
\begin{itemize}
    \item \textbf{Speech}: speaker gender, age, role, language, emotion, style, prosody, source.
    \item \textbf{SFX}: source, action/content, intensity, spatial property, temporal pattern, material, style.
    \item \textbf{BGM}: genre, mood, instrument, style, intensity, temporal pattern, spatial property.
    \item \textbf{Scene}: spatial property, temporal pattern, intensity, mood, style.
\end{itemize}

\paragraph{Question rewriting.}
Structured targets are rewritten into natural questions using an LLM with strict constraints. The model is instructed to preserve the \texttt{qa\_id}, \texttt{qa\_type}, event ID, modality, and attribute field; not to add or remove targets; and not to include answer choices in the question. The prompt also prevents exposing internal annotation terms such as \texttt{actor\_or\_source} or \texttt{action\_or\_content}. For \texttt{speech\_content}, the rewritten question must quote the exact spoken content or a long exact phrase.

The core rewriting instruction is:

\begin{lstlisting}
Rewrite each structured QA target as one natural yes/no question for evaluating a generated audio clip.

Requirements:
- Answerable solely from the audio.
- Preserve the original target exactly.
- Do not add or omit information.
- Do not expose internal field names.
- Keep the question concise and natural.

Return strict JSON:
{
  "questions": [
    {
      "qa_id": "same id as input",
      "question": "natural yes/no question"
    }
  ]
}
\end{lstlisting}

\paragraph{Leakage prevention and fallback templates.}
To reduce question leakage, we apply rule-based checks after rewriting. A question is replaced by a safe template if it copies the full event text into an attribute question, repeats the attribute value excessively, exposes internal field names, or contains forbidden words such as \texttt{reference}, \texttt{annotation}, \texttt{caption}, or \texttt{metadata}. We use component-aware fallback templates, \textit{e.g.,}
\begin{lstlisting}
    Event presence:  Is there a sound event described as {event_text}?
BGM genre:       Is there {value} background music in the audio?
SFX material:    Does the sound have a {value} quality?
Speech language: Is the speech in {value}?
Speech content:  Does the speaker say "{spoken_content}" in the audio?
\end{lstlisting}

\paragraph{Validation and filtering.}
Finally, each generated question is validated against its structured evidence. The validation checks whether the question refers to the same event, modality, and attribute, whether the attribute value is correctly represented, and whether unsupported details are introduced. Questions with severe validation errors are discarded. We also remove duplicated questions, questions with internal field wording, invalid \texttt{Scene} event-attribute questions, and malformed questions. After filtering, the final benchmark contains \textbf{25,707} rubrics pairs over \textbf{2,258} audio-text instances.





\section{Detailed Metric Definitions}
\label{app:metric_definitions}

This section provides the precise definitions of the fine-grained metrics used in AudioScape-TTA. Unless otherwise stated, all reported values are multiplied by $100$ and presented as percentages.

\paragraph{Semantic Satisfaction Rate.}
For each event-presence or event-attribute rubric $i$, let
$p_i\in\{\mathrm{Yes},\mathrm{No}\}$ denote the semantic answer predicted by
Qwen3-Omni-Instruct and let $y_i=\mathrm{Yes}$ denote the positive target
requirement derived from the annotation. Its binary satisfaction indicator is
$$
s_i=\mathbb{I}[p_i=y_i].
$$

For a speech-content rubric $i$, the satisfaction indicator is defined using
the ASR-based coverage criterion:
$$
s_i=\mathbb{I}[R_{\mathrm{mixed},i}\geq0.60].
$$

For any rubric set $\mathcal{S}$, the Semantic Satisfaction Rate is the
micro-average of item-level indicators:
\begin{align}
\mathrm{SR}(\mathcal{S})
&=
\frac{1}{|\mathcal{S}|}
\sum_{i\in\mathcal{S}} s_i, \\
\mathrm{Overall\ SR}
&=
\mathrm{SR}(\mathcal{Q}), \qquad
\mathrm{Event\ SR}
=
\mathrm{SR}(\mathcal{Q}_{\mathrm{event}}), \nonumber\\
\mathrm{Attribute\ SR}
&=
\mathrm{SR}(\mathcal{Q}_{\mathrm{attr}}).
\end{align}

\paragraph{Evaluator confidence.}
For event-presence and event-attribute rubrics, we additionally report an uncalibrated evaluator-confidence diagnostic. Let $\ell_A$ and $\ell_B$ be the log-probabilities assigned to the two answer options. We normalize them over the binary choice set:
\begin{equation}
q_A = \frac{\exp(\ell_A)}
{\exp(\ell_A)+\exp(\ell_B)}, \quad
q_B = \frac{\exp(\ell_B)}
{\exp(\ell_A)+\exp(\ell_B)}.
\end{equation}
and the reported value is the mean confidence over the corresponding rubrics subset. This quantity measures the evaluator's relative preference between the two candidate answers only. It is not calibrated as a probability of semantic correctness, is not an audio-quality metric, and should not be compared directly across different evaluator models.

\paragraph{Speech-content preprocessing.}
Speech content is evaluated by comparing the Qwen3-ASR transcript with the annotated target text. Before scoring, we remove a leading bracketed tag when present, lowercase Latin text, and collapse consecutive whitespace. Punctuation does not form an evaluation unit. We retain numbers, repeated words, and filler expressions; no stop-word removal, duplicate suppression, or prefix-based shortcut is applied.

We extract two unit types from the normalized text:
\begin{itemize}
    \item \textbf{Word units}: lowercased Latin alphanumeric tokens, optionally containing internal hyphens or apostrophes;
    \item \textbf{Character units}: individual Chinese CJK characters.
\end{itemize}
This mixed-unit formulation supports English, Chinese, and code-switched speech under a single protocol.

\paragraph{Mixed-unit coverage.}
Let $r$ and $\hat{r}$ denote the normalized reference text and ASR transcript, respectively. Let $c_r^w(u)$ and $c_{\hat{r}}^w(u)$ be the occurrence counts of word unit $u$ in the reference and transcript; $c_r^c(v)$ and $c_{\hat{r}}^c(v)$ are defined analogously for Chinese character unit $v$. We use clipped-count matching:
\begin{equation}
m_w = \sum_{u}\min\left(c_r^w(u), c_{\hat{r}}^w(u)\right),
\end{equation}
\begin{equation}
m_c = \sum_{v}\min\left(c_r^c(v), c_{\hat{r}}^c(v)\right).
\end{equation}
The total number of matched units is $m=m_w+m_c$. Let $N_r$ denote the total number of word and character units in the reference text. We define mixed-unit coverage as
\begin{equation}
R_{\mathrm{mixed}} = \frac{m}{N_r}.
\end{equation}
Clipped-count matching prevents a repeated transcript unit from matching
multiple occurrences in the reference. Importantly, this score is an
order-agnostic reference-unit coverage measure: it does not require the
matched units to occur contiguously or in the same order as in the
reference, and it does not penalize additional transcript units. This
coverage score is used only to determine SCCA@0.60; mixed-unit precision
and F1 are not reported as benchmark metrics.


\paragraph{Speech Content satisfaction rate.}
Our primary speech metric is Speech Content satisfaction rate at a mixed-unit coverage threshold $\tau$, denoted as SCCA@$\tau$. A speech-content item is counted as correct if its mixed-unit coverage reaches the threshold:

\begin{equation}
s_i^{\mathrm{speech}} =
\mathbf{1}\left(R_{\mathrm{mixed},i} \geq \tau\right).
\end{equation}
For the $N_{\mathrm{speech}}$ speech-content items, we compute
\begin{equation}
\mathrm{SCCA@}\tau =
\frac{1}{N_{\mathrm{speech}}}
\sum_{i=1}^{N_{\mathrm{speech}}}
\mathbf{1}\left(R_{\mathrm{mixed},i} \geq \tau\right).
\end{equation}

We use $\tau=0.60$ throughout the paper. Consequently, SCCA@0.60 measures whether the transcript contains clipped
matches for at least $60\%$ of the reference units under the mixed-unit
coverage criterion. It is not an order-sensitive exact-transcription
metric. The score pools clipped matches over Latin word units and Chinese
character units; it does not use a maximum across unit types or any
fixed-length prefix shortcut.

\paragraph{Supplementary WER and CER diagnostics.}
WER and CER are computed from the same Qwen3-ASR transcripts used for
speech-content evaluation. In contrast to SCCA@0.60, which is a rubric-level
binary target-coverage metric over $N_{\mathrm{speech}}=163$ speech-content
rubrics, WER and CER are clip-level corpus transcription diagnostics.

Consistent with the main-text statistics, the final benchmark contains 163
speech-content rubrics associated with 83 clips containing target utterances.
For the supplementary clip-level transcription diagnostics, we construct a
caption-derived reference transcript by concatenating quoted
\texttt{[Speaker ...]} utterances in annotation order. This construction is
available for 80 of the 83 target-utterance clips. The remaining three clips
contain speech-related targets without a usable quoted utterance reference
(e.g., indistinct or non-lexical vocal activity) and are excluded only from
the WER/CER computation; they remain included in the speech-content benchmark
and in SCCA@0.60.

In the final evaluation cache, WER is computed over
$|\mathcal{J}_{\mathrm{WER}}|=69$ valid reference clips containing a total of
$1{,}503$ Latin word units, and CER is computed over
$|\mathcal{J}_{\mathrm{CER}}|=11$ valid reference clips containing a total of
$309$ Chinese CJK character units. A code-switched clip may contribute to both
sets when it contains both unit types. Because CER is computed on a smaller
Chinese-reference subset, it should be interpreted as a complementary
diagnostic with additional caution. The resulting WER and CER values are
reported in Table~\ref{tab:speech_appendix}.

\paragraph{Modality, composition, and complexity subsets.}
For modality-level analysis, rubrics items are grouped according to the queried modality, including scene, sound effects, background music, and speech. In particular, \emph{Speech-modality rubrics satisfaction rate} denotes rubrics satisfaction rate over rubrics associated with the speech modality; it is distinct from SCCA@0.60, which evaluates the linguistic content of synthesized speech through ASR transcripts.

For composition analysis, samples are grouped by the set of modalities required by their prompts, such as Sound-only, Sound+Music, or Sound+Speech+Music. For complexity analysis, samples are grouped into Easy, Medium, and Hard subsets according to the benchmark's event-density and structural-complexity annotations. All subset results use the same item-level correctness definition as the overall rubrics-based evaluation.

\paragraph{Complementary global metrics.}
We additionally report global quality and alignment metrics as complementary measurements. Audiobox-Aesthetic provides content enjoyment (CE), content usefulness (CU), production complexity (PC), and production quality (PQ), where higher values indicate better predicted perceptual quality. $\mathbf{CLAP}_{\mathrm{MS}}$ measures global text--audio embedding similarity, where higher values indicate stronger coarse semantic alignment. Distribution-based metrics include Fr\'echet distances computed in pretrained audio-feature spaces, Kullback--Leibler divergence (KL), and Inception Score (ISC). Lower Fr\'echet distance and KL are preferred, whereas higher ISC is preferred. These global metrics are not included in SR and are intended to complement, rather than replace, the proposed fine-grained semantic evaluation.

















\begin{table*}[t]
\centering
\footnotesize
\setlength{\tabcolsep}{3.2pt}
\renewcommand{\arraystretch}{1.05}
\begin{tabular}{l|l|l|l|l}
\toprule
\textbf{Model}
& \textbf{Paradigm}
& \textbf{Text Encoder}
& \textbf{Audio Representation}
& \textbf{Generation Backbone} \\
\midrule

AudioLDM 2
& LD
& CLAP + FLAN-T5
& AudioMAE + Mel VAE
& UNet \\

Make-An-Audio 2
& LD
& CLAP + T5
& Mel VAE
& Transformer \\

AudioGen
& Autoregressive
& T5 Conditioner
& EnCodec Tokenizer
& AR Transformer \\

Tango~2
& LD + DPO
& FLAN-T5
& AudioLDM VAE
& UNet \\

TangoFlux
& Flow Matching
& FLAN-T5
& Stable Audio VAE
& Flux DiT \\

EzAudio
& DiT
& FLAN-T5
& Stable Audio VAE + DAC
& EzAudio-DiT \\

MAGNeT
& Masked Modeling
& T5-Large
& EnCodec Tokenizer
& Non-AR Transformer \\

Stable Audio Open
& DiT
& T5-Base
& Waveform VAE
& Diffusion Transformer \\

MMAudio
& Flow Matching
& CLIP
& Mel VAE + BigVGAN
& MM-DiT \\

Foley-Omni
& MM Flow Matching
& UM-T5
& Mel VAE + BigVGAN
& Conditional DiT \\

Omni2Sound
& Unified Diffusion
& FLAN-T5
& Wav VAE
& DiT \\

AudioStory
& LLM-guided Diffusion
& Qwen2.5 + T5
& Whisper + VAE
& LLM Planner + DiT \\

Dasheng AudioGen
& Unified Non-AR
& FLAN-T5-Large
& Dasheng Tokenizer
& Flow DiT \\

\bottomrule
\end{tabular}

\caption{
Architectural overview of the 13 open-source text-to-audio (TTA) models evaluated in AudioScape-TTA.
We summarize the generation paradigms, text encoders, audio representations, and generation backbones.
LD denotes Latent Diffusion, DPO denotes Diffusion Preference Optimization, 
DiT denotes Diffusion Transformer, AR denotes Autoregressive, and Non-AR denotes Non-Autoregressive generation.
For multimodal systems, only components relevant to text-to-audio generation are listed.
}
\label{tab:model_architectures}
\end{table*}

\section{Human Evaluation Protocol}
\label{app:human_protocol}

To validate the effectiveness of our automatic evaluation framework, we
conduct a human evaluation study on a subset of AudioScape-TTA. We randomly
select 10 prompts covering different modality compositions and complexity
levels, and collect generated audio from all 13 evaluated TTA models,
resulting in 130 audio samples. The identities of the models are anonymized,
and all samples are randomly shuffled before evaluation.

Each audio sample is independently rated by 26 participants. For each sample,
participants evaluate three aspects: \textit{event completeness},
\textit{attribute satisfaction}, and \textit{audio quality}. Event completeness
measures whether the required sound events described in the prompt are present
in the generated audio. Attribute satisfaction rate evaluates whether fine-grained
acoustic properties, such as sound source characteristics, temporal patterns,
and other specified attributes, are correctly realized. Audio quality measures
overall perceptual quality, including naturalness, coherence, and generation
artifacts.

Participants are provided with the original text prompt and the generated
audio, and rate each dimension independently using a 5-point Likert scale.
For semantic dimensions (\textit{event completeness} and
\textit{attribute satisfaction rate}), a higher score indicates that the generated
audio better satisfies the corresponding textual requirements. For
\textit{audio quality}, a higher score indicates better perceptual quality
and fewer noticeable artifacts. The final human score for each sample is
computed by averaging the ratings from all participants. These human
judgments are then used to analyze the correlation between human perception
and automatic evaluation metrics.

\begin{table*}[t]
\centering
\footnotesize
\setlength{\tabcolsep}{3.8pt}
\begin{tabular}{lccccccc}
\toprule
\textbf{Model}
& $\mathbf{FD}_{\mathrm{VGG}}\downarrow$
& $\mathbf{FD}_{\mathrm{PANN}}\downarrow$
& $\mathbf{FD}_{\mathrm{PASST}}\downarrow$
& $\mathbf{KL}_{\mathrm{PANN}}\downarrow$
& $\mathbf{KL}_{\mathrm{PaSST}}\downarrow$
& $\mathbf{ISC}_{\mathrm{PANN}}\uparrow$
& $\mathbf{ISC}_{\mathrm{PASST}}\uparrow$ \\
\midrule

AudioLDM 2
& 4.816 & 15.342 & 229.479 & 1.787 & 1.694 & 8.157 & 7.600 \\

Make-An-Audio 2
& 5.580 & 21.137 & 245.892 & 2.477 & 2.432 & 5.481 & 5.900 \\

AudioGen
& 3.976 & \textbf{2.874} & 155.277 & 2.064 & 1.930 & 7.255 & 7.648 \\

Tango 2
& 3.965 & 13.619 & 227.666 & 1.919 & 1.625 & 7.194 & 6.306 \\

TangoFlux
& 4.504 & 19.405 & 266.794 & 1.990 & 1.755 & 8.344 & 7.283 \\

EzAudio
& \underline{2.514} & \underline{9.711} & \textbf{119.163} & 1.851 & 1.740 & 8.291 & 8.515 \\

MAGNeT
& 15.711 & 38.295 & 509.685 & 4.015 & 3.324 & 1.913 & 1.736 \\

Stable Audio Open
& 3.114 & 14.089 & 199.270 & 2.035 & 1.912 & 9.418 & \textbf{8.678} \\

MMAudio
& 3.986 & 11.574 & 181.498 & 2.050 & 1.943 & 8.088 & 8.160 \\

Foley-Omni
& 2.728 & 16.391 & 211.450 & \textbf{1.628} & \underline{1.522} & 8.336 & 8.475 \\

Omni2Sound
& \textbf{1.620} & 10.160 & \underline{132.897} & \underline{1.773} & \textbf{1.520} & \textbf{10.330} & \underline{8.600} \\

AudioStory
& 4.394 & 21.132 & 308.961 & 2.309 & 2.183 & 8.379 & 7.607 \\

Dasheng AudioGen
& 3.174 & 17.661 & 202.953 & 1.948 & 1.737 & \underline{9.527} & 8.323 \\

\bottomrule
\end{tabular}
\caption{
Distribution-based audio quality metrics of open-source TTA models on
AudioScape-TTA.
These metrics evaluate global distributional similarity and perceptual quality
in pretrained feature spaces and are reported as complementary analyses rather
than components of the proposed semantic evaluation framework.
The best and second-best results are highlighted in \textbf{bold} and
\underline{underline}, respectively.
}
\label{tab:audio_quality}
\end{table*}

\begin{table*}[t]
\centering
\footnotesize
\setlength{\tabcolsep}{3.8pt}
\begin{tabular}{lcccccc}
\toprule
\textbf{Model} &
\textbf{Sound} &
\textbf{Music} &
\textbf{Sound+Speech} &
\textbf{Sound+Music} &
\textbf{Speech+Music} &
\textbf{Sound+Speech+Music} \\
\midrule

AudioLDM 2
& 69.19 & 84.93 & 52.37 & 70.54 & 56.25 & 48.60 \\

Make-An-Audio 2
& 65.65 & 72.33 & 50.50 & 54.44 & 56.25 & 47.18 \\

AudioGen
& 67.86 & 74.25 & 53.78 & 71.25 & 62.50 & 54.35 \\

Tango 2
& 79.53 & 80.27 & 60.29 & 77.93 & \underline{75.00} & 65.46 \\

TangoFlux
& \textbf{81.42} & 76.16 & 65.61 & 72.12 & 62.50 & 56.70 \\

EzAudio
& \underline{80.22} & \textbf{87.40} & 63.11 & \underline{81.08} & \underline{75.00} & 66.28 \\

MAGNeT
& 51.96 & \underline{86.85} & 26.47 & 73.05 & 50.00 & 39.90 \\

Stable Audio Open
& 70.01 & 73.42 & 48.57 & 65.13 & 56.25 & 46.91 \\

MMAudio
& 67.36 & 77.53 & 48.46 & 68.08 & 43.75 & 46.63 \\

Foley-Omni
& 80.63 & 84.93 & \underline{69.36} & \textbf{82.54} & \textbf{87.50} & \underline{72.69} \\

Omni2Sound
& 78.51 & 84.66 & 66.08 & 80.70 & \underline{75.00} & 71.98 \\

AudioStory
& 70.73 & 72.02 & 62.12 & 69.90 & 43.75 & 57.91 \\

Dasheng AudioGen
& 74.22 & 84.93 & \textbf{70.71} & 76.20 & \underline{75.00} & \textbf{74.88} \\

\bottomrule
\end{tabular}
\caption{Semantic Satisfaction Rate (SR, \%) of open-source text-to-audio generation models across samples with different modality compositions. The best and second-best results are highlighted in \textbf{bold} and \underline{underline}, respectively.}
\label{tab:modality_composition}
\end{table*}

\section{TTA Model Generation Protocol}

To ensure a model-native comparison across text-to-audio generation
models, we follow the native text-conditioning interface of each model and
report the input format, output configuration, and sampling settings. Unless
otherwise specified, each model generates one audio sample per prompt without
manual resampling, manual selection, or best-of output picking.

\subsection{Evaluation Input and Runtime Setup}

All evaluated models receive the text descriptions provided by AudioScape-TTA.
The input CSV contains the fields \texttt{name}, \texttt{caption}, and
\texttt{struct\_caption}. The \texttt{caption} field contains the free-form
natural language description and serves as the default generation prompt for
most evaluated models.
Most models directly use the free-form \texttt{caption} field as input.
For models supporting structured prompting, i.e., Make-An-Audio-2 and
Dasheng-AudioGen, the annotated \texttt{struct\_caption} is converted into
the corresponding model-specific prompt representation.

All models are inferred on NVIDIA H20 GPUs with CUDA Version 12.5.

\subsection{Per-Model Generation Protocol}

We use the official open-source checkpoints and inference pipelines of all
evaluated TTA models. Unless otherwise specified, each model generates one
10-second audio sample per input prompt without manual selection or
resampling. 
The reported configurations are collected from official checkpoints,
released inference scripts, and configuration files. 

\vspace{0.4em}
\noindent\textbf{AudioLDM2}\footnote{\url{https://huggingface.co/cvssp/audioldm2}.}
 We use 200 diffusion steps
with a 10-second generation length.

\vspace{0.4em}
\noindent\textbf{Make-An-Audio-2}\footnote{\url{https://huggingface.co/ByteDance/Make-An-Audio-2}.}
Generation uses DDIM sampling with 100 steps and guidance scale 4.

\vspace{0.4em}
\noindent\textbf{AudioGen}\footnote{\url{https://huggingface.co/facebook/audiogen-medium}.}
The text caption is directly used with the default generation
configuration of the official checkpoint.

\vspace{0.4em}
\noindent\textbf{Tango~2}\footnote{\url{https://huggingface.co/declare-lab/tango2}.}
We use the default text-to-audio pipeline with 200 sampling steps and
guidance scale 3.

\vspace{0.4em}
\noindent\textbf{TangoFlux}\footnote{\url{https://huggingface.co/declare-lab/TangoFlux}.}
Generation uses the default flow-matching configuration with 50 steps and
guidance scale 4.5.

\vspace{0.4em}
\noindent\textbf{EzAudio}\footnote{\url{https://huggingface.co/OpenSound/EzAudio}.}
We use the official inference configuration with 100 DDIM steps and guidance
scale 5.

\vspace{0.4em}
\noindent\textbf{MAGNeT}\footnote{\url{https://huggingface.co/facebook/magnet-medium-10secs}.}
The official 10-second checkpoint is used with its default generation
configuration.

\vspace{0.4em}
\noindent\textbf{Stable Audio Open}\footnote{\url{https://huggingface.co/stabilityai/stable-audio-open-1.0}.}
We use the official inference configuration with 100 sampling steps and
classifier-free guidance scale 7.

\vspace{0.4em}
\noindent\textbf{MMAudio}\footnote{\url{https://huggingface.co/hkchengrex/MMAudio}.}
 The model generates audio using
25 sampling steps and guidance strength 4.5.

\vspace{0.4em}
\noindent\textbf{Foley-Omni}\footnote{\url{https://huggingface.co/CocoBro/Foley-Omni}.}
We use the default text-to-audio setting with 50 sampling steps and audio
guidance scale 3.

\vspace{0.4em}
\noindent\textbf{Omni2Sound}\footnote{\url{https://huggingface.co/Dalision/Omni2Sound}.}
The caption is converted into the required text prompt field. We use the
default configuration with 100 steps and guidance scale 2.5.

\vspace{0.4em}
\noindent\textbf{AudioStory}\footnote{\url{https://huggingface.co/TencentARC/AudioStory-3B}.}
Generation
uses 50 inference steps with guidance scale 4.

\vspace{0.4em}
\noindent\textbf{Dasheng-AudioGen}\footnote{\url{https://huggingface.co/mispeech/Dasheng-AudioGen}.}
We use the official inference pipeline.

\subsection{Model Architecture and Design Observations}

Although TTA performance depends on multiple factors, including training data,
model scale, text supervision, and inference configuration, our benchmark
reveals several architectural and design tendencies among current
open-source models.

First, recent unified and diffusion-transformer (DiT)-based generation models
tend to achieve stronger overall performance in our evaluation compared with
earlier autoregressive or masked-token systems. However, this observation may
also be influenced by differences in training data, model scale, and
optimization strategies. For example, Foley-Omni achieves the best overall
performance and remains robust under hard and multi-modal settings, suggesting
that unified modeling across speech, sound effects, and music can be
advantageous for structured soundscape generation. Similarly, EzAudio and
Omni2Sound rank among the top-performing systems, indicating that recent
unified generation paradigms provide strong semantic coverage and compositional
capability.

Second, latent diffusion and flow-matching models generally achieve competitive
event-level alignment, but their fine-grained attribute fidelity varies
substantially. Tango~2 and TangoFlux obtain strong overall performance,
reflecting the strong text--audio alignment capability of these diffusion-based
generation paradigms. However, their performance decreases more noticeably on
hard and structurally complex samples, indicating that strong global prompt
following does not necessarily guarantee reliable control over detailed
acoustic attributes. This observation is consistent with our attribute-level
evaluation, where current TTA systems still struggle with source identity,
material characteristics, temporal dynamics, spatial relations, and
speaker-related attributes.

Third, speech-content generation remains a distinct challenge compared with
general sound-event synthesis. Dasheng AudioGen achieves substantially better
performance on speech-related evaluation, which is consistent with its
speech-oriented modeling design and dedicated audio representation. In contrast,
many general-purpose TTA models can generate speech-like acoustic patterns but
fail to preserve verifiable linguistic content, highlighting the need for
explicit speech-content modeling rather than treating speech as another generic
audio event.

Finally, the comparison between rubrics-based evaluation and conventional global
similarity metrics demonstrates that strong global text--audio similarity does
not necessarily imply fine-grained semantic correctness. Some models achieve
competitive CLAP-based alignment scores but perform noticeably worse under
event- and attribute-level rubrics-based evaluation. This indicates that optimizing for
perceptual plausibility or coarse text--audio alignment alone is insufficient
to satisfy structured semantic requirements. Overall, our findings suggest
that future TTA systems should combine high-quality acoustic generation with
stronger mechanisms for explicit event grounding, fine-grained attribute
control, reliable speech-content preservation, and robust multi-modal
composition.

\begin{table}[t]
\centering
\small
\setlength{\tabcolsep}{6pt}
\begin{tabular}{lcc}
\toprule
\textbf{Model} &
\textbf{WER}$\downarrow$ &
\textbf{CER}$\downarrow$ \\
\midrule
AudioLDM 2         & 1.1949 & 1.0000 \\
Make-An-Audio 2    & 1.0093 & 1.0000 \\
AudioGen          & 1.1284 & 1.0000 \\
Tango 2            & 1.2448 & 1.1294 \\
TangoFlux         & 1.1756 & 0.9968 \\
EzAudio           & 1.2582 & 0.9968 \\
MAGNeT            & 1.8603 & 1.0000 \\
Stable Audio Open & 1.0745 & 1.0712 \\
MMAudio           & 1.0153 & 1.0000 \\
Foley-Omni        & \underline{0.5469} & \textbf{0.9191} \\
Omni2Sound        & 1.0253 & 1.0000 \\
AudioStory        & 1.1564 & 1.0356 \\
Dasheng AudioGen  & \textbf{0.2395} & \underline{0.9935} \\
\bottomrule
\end{tabular}

\caption{
Supplementary clip-level ASR transcription diagnostics.
WER is computed over 69 valid reference clips containing 1,503 Latin word
units, while CER is computed over 11 valid reference clips containing 309
Chinese CJK character units. Both metrics are computed from Qwen3-ASR
transcripts of generated audio and caption-derived target utterances.
They are complementary diagnostics and do not alter the 163-rubric
SCCA@0.60 benchmark evaluation.
The best and second-best results are highlighted in \textbf{bold} and
\underline{underline}, respectively.
}
\label{tab:speech_appendix}

\end{table}

\begin{table*}[t]
\centering
\small
\setlength{\tabcolsep}{5pt}
\begin{tabular}{lccccc}
\toprule
\textbf{Model} & $\mathrm{SCCA}@0.40$ & $\mathrm{SCCA}@0.50$ &
$\mathrm{SCCA}@0.60$ & $\mathrm{SCCA}@0.70$ & $\mathrm{SCCA}@0.80$ \\
\midrule
AudioLDM 2        & 1.23 & 1.23 & 0.00 & 0.00 & 0.00 \\
Make-An-Audio 2   & 1.23 & 1.23 & 0.00 & 0.00 & 0.00 \\
AudioGen          & 0.00 & 0.00 & 0.00 & 0.00 & 0.00 \\
Tango 2           & 0.00 & 0.00 & 0.00 & 0.00 & 0.00 \\
TangoFlux         & 0.00 & 0.00 & 0.00 & 0.00 & 0.00 \\
EzAudio           & 0.00 & 0.00 & 0.00 & 0.00 & 0.00 \\
MAGNeT            & 0.00 & 0.00 & 0.00 & 0.00 & 0.00 \\
Stable Audio Open & 0.00 & 0.00 & 0.00 & 0.00 & 0.00 \\
MMAudio           & 0.00 & 0.00 & 0.00 & 0.00 & 0.00 \\
Foley-Omni        & \underline{68.10} & \underline{66.26} & \underline{57.06} & \underline{50.92} & \underline{45.40} \\
Omni2Sound        & 0.00 & 0.00 & 0.00 & 0.00 & 0.00 \\
AudioStory        & 0.00 & 0.00 & 0.00 & 0.00 & 0.00 \\
Dasheng AudioGen  & \textbf{80.98} & \textbf{80.37} & \textbf{77.30} & \textbf{72.39} & \textbf{68.71} \\
\bottomrule
\end{tabular}
\caption{Speech-Content Coverage Satisfaction Rate (\%) at Different Thresholds under different SCCA
thresholds. Each value is the percentage of 163 speech-content rubrics whose
mixed-unit target coverage is at least $\tau$.}
\label{tab:scca_threshold_sensitivity}
\end{table*}

\section{Additional Experimental Results}

\subsection{Modality-combination Analysis}
In the main paper, we analyze modality-level performance by grouping rubrics pairs according to the queried modality, i.e., \emph{sound}, \emph{music}, and \emph{speech}. This reveals the modality-specific strengths of different models. In contrast, here we group each sample according to the modality composition required by its prompt. Therefore, this analysis evaluates \emph{compositional robustness} rather than isolated modality-specific capability.
The results are shown in Table~\ref{tab:modality_composition}.

The results show that strong single-modality performance does not always transfer to multi-modal settings. TangoFlux achieves the highest satisfaction rate on \emph{Sound}-only samples (\textbf{81.42\%}), but drops substantially on \emph{Sound+Speech+Music} samples (56.70\%). Similarly, MAGNeT performs competitively on \emph{Music}-only samples (\underline{86.85\%}), yet struggles on \emph{Sound+Speech} (26.47\%) and \emph{Sound+Speech+Music} (39.90\%). These results suggest that generating isolated sound or music events is insufficient for handling complex prompts where multiple modalities co-occur.

For mixed \emph{Sound+Music} scenes, Foley-Omni achieves the best satisfaction rate (\textbf{82.54\%}), outperforming EzAudio (\underline{81.08\%}) and Omni2Sound (80.70\%). This complements the main-paper findings: Foley-Omni is not only strong on sound-effect questions, but also better at coordinating sound events with background music within the same scene. For speech-involved compositions, Dasheng AudioGen performs best on both \emph{Sound+Speech} (\textbf{70.71\%}) and \emph{Sound+Speech+Music} (\textbf{74.88\%}), consistent with its superior speech-content satisfaction rate reported in the main paper. This indicates that speech generation becomes a key bottleneck when speech is embedded in complex multi-modal scenes.

Overall, this analysis demonstrates that modality-composition evaluation provides complementary insights beyond conventional modality-wise evaluation. 
Foley-Omni exhibits strong compositional robustness across different modality combinations, EzAudio remains highly competitive in music-involved scenarios, and Dasheng AudioGen is particularly effective on speech-involved compositions. These findings further highlight the importance of fine-grained compositional evaluation for text-to-audio generation systems.

\subsection{Audio Quality Evaluation}

In addition to semantic requirement satisfaction, we report conventional
perceptual and distribution-based metrics as complementary analyses.
Specifically, Audiobox-Aesthetic evaluates perceptual quality from multiple
dimensions, including clarity, coherence, production complexity, and overall
quality. Distribution-based metrics, including Fréchet distance (FD),
KL divergence, and Inception Score (ISC), measure the similarity and diversity
of generated audio feature distributions.

Table~\ref{tab:audiobox_aesthetic} and Table~\ref{tab:audio_quality}
summarize these complementary evaluations. The rankings obtained from these
metrics are not always consistent with semantic requirement satisfaction.
For example, models achieving strong distributional similarity or perceptual
scores do not necessarily achieve the highest semantic satisfaction rates.
This difference is expected because these metrics mainly characterize global
audio quality, realism, or feature-level distribution matching, while our
rubrics-based evaluation directly verifies whether prompt-specified semantic
requirements are realized.

Therefore, these metrics are reported to provide a broader view of model
performance, but are not used as the primary criteria for evaluating
structured soundscape instruction following.
\subsection{Additional Speech Content Evaluation}

\paragraph{Additional Speech-Content Diagnostics.} Besides the binary speech-content satisfaction rate reported in the main paper using
SCCA@0.60, we additionally report Word Error Rate (WER) and Character Error
Rate (CER) in Table~\ref{tab:speech_appendix}. These metrics are computed only
for samples containing speech by comparing ASR transcriptions against the
reference text.

Unlike SCCA@0.60, which measures whether the required speech content is
sufficiently covered for semantic evaluation, WER and CER quantify full
transcription similarity and are therefore more sensitive to insertion,
deletion, and substitution errors. Since many evaluated open-source TTA models
are not explicitly designed for controllable speech generation, WER/CER are
reported as complementary analyses rather than used in the benchmark's primary
semantic ranking.

\paragraph{Speech-content coverage threshold sensitivity.}
Table~\ref{tab:scca_threshold_sensitivity} shows that the central speech
conclusion is not driven by the choice of $\tau=0.60$. Across the stricter
range $\tau\in\{0.60,0.70,0.80\}$, Dasheng AudioGen consistently obtains the
highest SCCA, while Foley-Omni consistently ranks second. 
Lowering the threshold to 0.40 or 0.50 yields only marginal non-zero scores (1.23\%) for AudioLDM 2 and Make-An-Audio 2, while preserving the same two leading models.

\begin{table}[t]
\centering
\small
\setlength{\tabcolsep}{5pt}
\begin{tabular}{lcc}
\toprule
\textbf{Comparison} & \textbf{Spearman $\rho$} & \textbf{Two-sided $p$} \\
\midrule

$\mathrm{SCCA}@0.60$ vs.\ $-\mathrm{WER}$ & 0.629 & 0.021 \\
\midrule
$\mathrm{SCCA}@0.60$ vs.\ $-\mathrm{CER}$ & 0.653 & 0.015 \\
\bottomrule
\end{tabular}
\caption{Coarse model-level rank agreement between SCCA@0.60 and complementary
clip-level edit-distance diagnostics across 13 TTA models.
}

\label{tab:scca_edit_distance_consistency}
\end{table}

\paragraph{Correlation between SCCA and WER.} 
Table~\ref{tab:scca_edit_distance_consistency} compares SCCA@0.60 with the
complementary ASR transcription diagnostics at the model level. We compute
Spearman correlation using average ranks for tied SCCA scores and correlate
SCCA@0.60 with $-\mathrm{WER}$ and $-\mathrm{CER}$, since lower edit error
indicates better transcription fidelity. SCCA@0.60 shows moderate positive
agreement with $-\mathrm{WER}$ ($\rho=0.629$, $p=0.021$) and
$-\mathrm{CER}$ ($\rho=0.653$, $p=0.015$). Because most general-purpose
models obtain tied zero SCCA@0.60 scores, these results should be interpreted
as coarse agreement between target-content coverage and edit-distance-based
transcription fidelity, rather than as a fine-grained ranking among
lower-performing systems.

\begin{table}[t]
\centering
\footnotesize
\setlength{\tabcolsep}{5pt}
\begin{tabular}{l|cccc}
\toprule
\textbf{Model}
& \textbf{CE}$\uparrow$
& \textbf{CU}$\uparrow$
& \textbf{PC}$\uparrow$
& \textbf{PQ}$\uparrow$ \\
\midrule

AudioLDM 2
& 4.21 & \underline{6.04} & 3.45 & \underline{6.34} \\

Make-An-Audio 2
& 3.37 & 5.14 & 3.15 & 5.67 \\

AudioGen
& 3.24 & 4.57 & 3.60 & 5.26 \\

Tango 2
& 4.13 & 5.61 & \underline{4.32} & 6.18 \\

TangoFlux
& \underline{4.27} & 5.53 & 4.04 & 6.02 \\

EzAudio
& 4.16 & 5.58 & 4.07 & 5.96 \\

MAGNeT
& 4.20 & 5.11 & \textbf{5.28} & 5.74 \\

Stable Audio Open
& 3.38 & \textbf{6.33} & 1.96 & \textbf{6.64} \\

MMAudio
& 3.79 & 5.25 & 3.33 & 5.77 \\

Foley-Omni
& 4.14 & 5.56 & 3.65 & 5.88 \\

Omni2Sound
& 4.21 & 5.80 & 3.95 & 6.24 \\

AudioStory
& 3.96 & 5.36 & 3.33 & 5.94 \\

Dasheng AudioGen
& \textbf{4.29} & 5.81 & 3.56 & 6.18 \\

\bottomrule
\end{tabular}

\caption{
Audiobox-Aesthetic evaluation results of open-source TTA models on
AudioScape-TTA.
CE, CU, PC, and PQ denote the aesthetic dimensions measured by Audiobox.
These metrics evaluate perceptual audio quality and are reported as
complementary analyses rather than semantic requirement evaluation.
The best and second-best results are highlighted in \textbf{bold} and
\underline{underline}, respectively.
}
\label{tab:audiobox_aesthetic}
\end{table}

\subsection{Sanity Checks for Audio-Grounded Rubric-Based Evaluation}

\paragraph{Audio-conditioned controls.}
To verify whether rubrics performance genuinely relies on acoustic evidence, we conduct two audio-conditioned control experiments using the reference-audio rubrics pairs. These experiments follow the same audio-conditioned setting as our evaluation protocol, where each rubrics item is evaluated together with an audio input. In the \textbf{audio-masked} setting, the original audio is replaced with a silent clip. In the \textbf{shuffled-audio} setting, each question is paired with the reference audio from a randomly selected different sample. As shown in Table~\ref{tab:qa_audio_control}, both control settings lead to a substantial drop in Overall SR, Event SR, and Attribute SR compared with the correctly matched reference-audio setting. SR follows the benchmark definition and includes event-presence, event-attribute, and thresholded speech-content correctness. These results demonstrate that high semantic satisfaction rates depend on informative and correctly matched
audio evidence rather than the question format alone.

\paragraph{Evaluator capability check.}
To investigate whether remaining errors on reference audio are partially caused
by evaluator limitations, we additionally evaluate the same QA pairs using
Gemini, a stronger audio-language model. As shown in
Table~\ref{tab:evaluator_capability_check}, Gemini-2.5-Pro achieves an Overall rubric
satisfaction rate of 95.03\% on the same rubric annotations and reference
audio, substantially outperforming Qwen3-Omni-Instruct.
The higher agreement obtained with Gemini-2.5-Pro is consistent with evaluator
capacity affecting reference-audio rubric scores, but does not definitively
separate evaluator errors from rubric-construction errors. Therefore, while we
adopt Qwen3-Omni-Instruct throughout the benchmark for accessibility and
reproducibility, the proposed evaluation protocol remains compatible with
stronger audio-language models and can benefit from future improvements in
audio reasoning capability.

\begin{table}[t]
\centering

\small
\setlength{\tabcolsep}{6pt}
\begin{tabular}{lccc}
\toprule
\textbf{Setting} &
\textbf{Overall SR} &
\textbf{Event SR} &
\textbf{Attribute SR} \\
\midrule
Reference Audio & \textbf{84.88} & \textbf{85.60} & \textbf{84.37} \\
Audio Masked    & 18.31 & 10.42 & 23.99 \\
Shuffled Audio  & 25.88 & 22.84 & 28.07 \\
\bottomrule
\end{tabular}
\caption{
Audio-conditioned control experiments using manipulated reference audio.
The results demonstrate that rubrics-based evaluation requires informative and correctly matched acoustic evidence.
}
\label{tab:qa_audio_control}
\end{table}

\begin{table}[t]
\centering
{\small
\setlength{\tabcolsep}{2.5pt}
\renewcommand{\arraystretch}{1.0}
\begin{tabular}{@{}lccc@{}}
\toprule
\textbf{Evaluator} &
\textbf{Overall SR} &
\textbf{Event SR} &
\textbf{Attribute SR} \\
\midrule
Qwen3-Omni-Instruct & 84.88 & 85.60 & 84.37 \\
Gemini-2.5-Pro      & 95.03 & 96.64 & 93.87 \\
\bottomrule
\end{tabular}
}
\caption{Evaluator capability check on the reference-audio QA pairs. Gemini-2.5-Pro is included as a stronger reference evaluator to to assess the effect of evaluator capacity on reference-audio rubric scores.}
\label{tab:evaluator_capability_check}
\end{table}

\subsection{Evaluator Bias and Answer-Position Analysis}
\label{app:evaluator_sensitivity}

Our main \textit{semantic satisfaction} metric measures whether
prompt-specified requirements are realized in generated audio. Since the
benchmark rubrics are formulated as positive requirement verification
questions, we additionally analyze two potential evaluator biases:
(i) a tendency to answer ``Yes'' regardless of audio evidence, and
(ii) sensitivity to the position of the affirmative answer. These analyses
evaluate evaluator reliability and do not extend the main metric into an
open-ended extra-content hallucination measure.

\paragraph{Balanced reference-audio sensitivity analysis.}
We construct a contrastive subset from reference audio by pairing each
event-presence or event-attribute positive rubric with a same-modality
contrastive target that is absent from the reference clip. To avoid exposing
the intended polarity to the evaluator, candidate rubrics are independently
verified using two Gemini-2.5-Pro prompts without providing the expected label,
candidate source, or polarity. We retain only complete positive-negative pairs
for which both items pass verification. The resulting held-out subset contains
452 semantic items from 226 complete pairs, with balanced labels
(226 ``Yes'' and 226 ``No'' items).

We evaluate \texttt{Qwen3-Omni-Instruct} under both answer orders for every
item (``Yes'' as option A and ``Yes'' as option B), resulting in 904 evaluator
calls. As shown in Table~\ref{tab:reference_sensitivity}, Qwen3-Omni-Instruct
achieves high discrimination capability on the reference-audio subset, with
94.25\% recall, 4.42\% false-positive rate, and 94.91\% balanced accuracy.
These results rule out a trivial always-``Yes'' answering strategy and indicate
limited answer-position sensitivity in this controlled reference-audio
analysis.

Importantly, this experiment is a \emph{Gemini-verified reference-audio
robustness analysis} rather than a human-annotated calibration set. Therefore,
the reported false-positive rate, precision, and calibration statistics should
be interpreted as evaluator sensitivity estimates with respect to the
verified subset, rather than definitive human-grounded error rates.

\begin{table}[t]
\centering
\small
\begin{tabular}{lc}
\toprule
\textbf{Metric} & \textbf{Result} \\
\midrule
Recall (TPR) & $94.25\%$ \\
False-positive rate (FPR) & $4.42\%$ \\
Specificity (TNR) & $95.58\%$ \\
Precision & $95.52\%$ \\
F1 score & $94.88\%$ \\
Accuracy & $94.91\%$ \\
Balanced accuracy & $94.91\%$ \\
ECE & $0.0197$ \\
Brier score & $0.0373$ \\
\bottomrule
\end{tabular}
\caption{
Evaluator sensitivity analysis of \texttt{Qwen3-Omni-Instruct} on the
Gemini-verified reference-audio subset.
TPR (recall) measures the fraction of verified satisfied requirements
correctly identified by the evaluator, while FPR measures the fraction of
verified unsatisfied requirements incorrectly predicted as satisfied.
TNR denotes specificity ($1-\mathrm{FPR}$). Precision and F1 quantify the
discrimination quality of affirmative predictions. ECE and Brier score measure
probability calibration.
The subset contains 452 balanced semantic items, including 226 positive and
226 negative items.
}
\label{tab:reference_sensitivity}
\end{table}

\paragraph{Answer-position sensitivity.}
We further randomize whether the affirmative answer appears as option A or
option B. On the balanced reference-audio subset, the balanced-accuracy gap
between the two answer orders is $1.33$ percentage points, and only
14 of 452 semantic items ($3.10\%$) receive different semantic predictions
under the two orders. This suggests limited answer-position sensitivity at
the item level in the controlled setting.

We further assess whether answer-order randomization affects the relative
comparison among evaluated TTA systems. As shown in
Table~\ref{tab:answer_order_rank_stability}, the fixed-order and
randomized-order model rankings exhibit strong agreement, with Spearman
correlations ranging from 0.937 to 0.993 across semantic dimensions. The
top-three models are identical under both protocols for overall, event-level,
and attribute-level semantic satisfaction. These results indicate that the
benchmark-level comparative conclusions are robust to answer-order
randomization.

\begin{table}[t]
\centering
\small
\begin{tabular}{lcc}
\toprule
\textbf{Metric} &
\textbf{Spearman $\rho$} &
\textbf{Top-3 Overlap} \\
\midrule
Overall Semantic Satisfaction & 0.979 & 3/3 \\
Event Semantic Satisfaction   & 0.937 & 3/3 \\
Attribute Semantic Satisfaction & 0.993 & 3/3 \\
\bottomrule
\end{tabular}
\caption{
Robustness of benchmark-level \emph{model rankings} to answer-order
randomization across 13 TTA systems. Spearman correlation is computed between
the fixed-order and randomized-order model rankings.
}
\label{tab:answer_order_rank_stability}
\end{table}

\paragraph{Scope of contrastive negatives.}
The contrastive negative rubrics above are used only to measure evaluator
presence--absence discrimination on reference audio. They are not used to
penalize generated audio for containing arbitrary additional sounds: a sound
absent from a reference clip may still be compatible with a generation prompt.
Accordingly, our semantic satisfaction metric measures satisfaction of
prompt-specified target content, rather than unrestricted extra-content
hallucination. Evaluating the latter would require prompts or annotations with
explicit \emph{forbidden} sound constraints, or dedicated human judgments.


\end{document}